\documentclass[prd,11pt]{revtex4-2}
\usepackage{graphicx}% Include figure files
\usepackage{dcolumn}% Align table columns on decimal point
\usepackage{bm}%
\usepackage{amsmath,latexsym,amssymb}
\usepackage{color}
\usepackage{latexsym}

\newcommand{\be}{\begin{equation}}
\newcommand{\ee}{\end{equation}}
\newcommand{\bea}{\begin{eqnarray}}
\newcommand{\ena}{\end{eqnarray}}

\newcommand{\nb}{\nonumber}

\newcommand{\ha}{\frac{1}{2}}
\newcommand{\plm}{M_{pl}}
\newcommand{\de}{\partial}
\newcommand{\ea}{\end{eqnarray}}

\begin{document}

  \title{Dynamical Cosmological Constant}
%\thanks{Footnote to title of article.}

  \author{G. di Donato}
  \author{L. Pilo}%
  \affiliation{Dipartimento di Scienze Fisiche e Chimiche, Universit\`a degli Studi dell'Aquila,  I-67100 L'Aquila, Italy}%
 \affiliation{ INFN, Laboratori Nazionali del Gran Sasso, I-67100 Assergi, Italy}%Lines break automatically or can be forced with \\

% \email{Luigi.Pilo@aquila.infn.it}

% \author{ Luigi Pilo}
%  \homepage{http://www.Second.institution.edu/~Charlie.Author.}
% \affiliation{%
% Second institution and/or address%\\This line break forced% with \\
% }%

\date{\today}% It is always \today, today,

\begin{abstract}
  The dynamical realisation of the equation  of state $p +\rho =0$ 
 is studied. A non-pathological dynamics for the perturbations of such
 a system mimicking  a  dynamical cosmological constant (DCC)
  requires to go beyond the perfect fluid paradigm. It is shown that
  an anisotropic stress must be always present.
The Hamiltonian of the system in isolation resembles the one of a
Pais-Uhlenbeck oscillator and linear stability requires that it cannot  be positive definite.
 The dynamics of linear cosmological perturbations in
  a DCC dominated Universe is studied in detail showing that when DCC
  is minimally coupled to gravity no dramatic instability is
  present. % {\underline{This  provides an example of a system where $H$ is not positive
  % definite but no fast instability occurs} \color{red} cancellare ripetizione ?}.
In contrast to what happens in  a cosmological  constant dominated Universe, 
the  non-relativistic matter contrast is no longer constant and exhibits an
oscillator behaviour at small scales while it grows weakly  at large
scales. In the gravitational waves sector, at small scales, the amplitude is  still suppressed as the inverse power of
the scale factor while it grows logarithmically at large scales. Also
the vector modes propagate, though no growing mode is found.
\end{abstract}

%Use showkeys class option if keyword
\maketitle

\section{Introduction}
We still do not know the nature of dark energy that is driving the
present acceleration of our Universe, recent observations (see for
instance~\cite{DES:2022fqx}) are consistent with the LCDM model that 
represents the simplest option. The motivations to go beyond a
cosmological constant are two fold: from a phenomenological point of
view, it is important to keep our options open  in the case
observations show any sizeable deviation from the ``vanilla'' scenario; from a theoretical perspective, it is rather challenging to
come up with a dynamical dark energy model. In a
Friedman-Robertson-Walker (FRW) geometry   to get into an
accelerated expansion regime, the strong energy condition (SEC) must be
violated  and the equation of state corresponding to a cosmological
constant already saturates the null energy condition (NEC),
 the weakest of all energy conditions. If one pushes
past NEC in the region $w<-1$, the scale factor explodes at a finite
time and generically small perturbations will trigger instabilities~\cite{Dubovsky:2005xd}. In the present paper we will focus on the
case $w=-1$, studying what are the constraints on a general
self-gravitating medium that saturates the NEC  ($w=-1$) in order that the dynamics
of its elementary excitations are healthy. It turns out that more degrees of
freedom of the ones present in a perfect fluid are needed. The
approach that we follow is to effectively describe the dark energy medium by
four scalar fields minimally coupled with gravity fluctuating around a non-trivial background. Such phonon-like fluctuations can be interpreted
as the Goldstone bosons for the  spontaneously broken translations; as a
consequence,  the low energy dynamics of the  fluctuations is dictated by
the symmetry breaking pattern. The bottom line is that,
 given the most general non-dissipative medium with $w=-1$, the
 requirement of a healthy dynamics for its elementary excitations 
selects a supersolid that will be  our model for a dynamical
cosmological constant (DCC). In a FRW Universe dominated by such a DCC, the dynamics of perturbation is rather different from
LCDM and will be studied in detail.

The outline of the paper is the following. In section \ref{de},
starting from the issues of k-essence, we study
the dynamical and stabilities properties of  a generic self-gravitating medium described in terms of
four scalar fields that corresponds to the four independent phonon-like
modes. Section \ref{cosmpert} is devoted to cosmological perturbations in Universe
dominated by a DCC. The impact of dark energy on
structure formation at linear order in perturbation theory is
discussed in section \ref{growth}. The propagation of gravitational waves is
described in section \ref{gravwav}, while section \ref{sect-vector} is
devoted to the study of vector modes. The conclusions are drawn in section \ref{concl}.
\section{Dynamical Cosmological Constant as a Self-gravitating Medium}
\label{de}
 In cosmology the description of matter as some sort of
 fluid has been rather successful and it is natural to  follow the
 same approach also for dark energy. In its simplest form, dark energy
 can be  defined as a component that, in the contest of a homogeneous
 FRW Universe, contributes with an energy momentum tensor (EMT) of a perfect fluid~\footnote{We use the signature $-+++$.}
 \be
 T_{\mu \nu} = ( p+\rho) u_\mu  u_\nu + p   \,
 g_{\mu \nu} ;
 \label{pfemt}
 \ee
where  $p$ is the  pressure and   $\rho$ is the energy density, such that $p= w \,
\rho$ with $w < - 1/3$, see~\cite{Kunz:2012aw,Amendola:2015ksp}
 for a recent discussion.
 In the case of a cosmological constant $\Lambda$: no additional degrees of
  freedom are present, $w$ is $-1$ and the EMT is proportional to the
 metric with $\rho=\Lambda$. Actually, if one sets $w=-1$ in (\ref{pfemt}), the conservation of  the EMT tensor gives
 automatically $\rho= \text{constant}$. In the following we will focus on the case
 $w=-1$, the most challenging to realize dynamically.

A better physical insight can be obtained by
 considering a generic k-essence~\cite{Chiba:1999ka,Armendariz-Picon:2000nqq} scalar field theory with 
 Lagrangian~\footnote{We only consider minimally coupled
   scalar fields. For the case a scalar tensor where the scalar field
   is non-minimally coupled to gravity, 
   solar system tests~\cite{Will:2014kxa} and the measurement of the
   propagation speed of gravitational waves~\cite{LIGOScientific:2017zic} put severe
   constraints~\cite{Lombriser:2016yzn,Ezquiaga:2017ekz,Creminelli:2017sry} on such a theories.}  $K(X, \, \Phi)$, where $X= - \ha g^{\mu \nu}
 \de_\mu \Phi \de_\nu \Phi$ and  with an EMT of
 the form (\ref{pfemt}).  In a standard   homogeneous and
   isotropic FRW cosmology, at the
 background level, denoting by $a$ the scale factor, the k-essence field has a profile of the form 
 \be
 \Phi =  \phi(t) \, , \qquad X=\bar X =
 \ha \dot{\phi}^2 \, ;
 \ee
 with
 \be
 \bar \rho = 2 \, \bar X \,   \bar K_X - \bar K \, , \qquad \bar
 p = \bar K \, ;
 \ee
where $t$ is the physical time, the bar stands for the background value,  while $K_X$, $K_\Phi$
 denote the partial derivative of $K$ with respect to $X$ and $\Phi$
 respectively. The time derivative with respect to
   physical time is denoted by a dot. Cosmological perturbations in FRW Universe dominated
 by k-essence are discussed in appendix \ref{app-kess}. Imposing  $w=-1$ gives
\be
 \bar X \,   \bar K_X  =0 \, .
\label{conseqs}
 \ee
Consider the following shift transformation
 \be
 \Phi \to \Phi+ \text{constant}  \, .
 \label{shiftsym}
 \ee
Unless $K$ is invariant under (\ref{shiftsym}) and then it depends
only on $X$, eq. (\ref{conseqs})
and the equation of motion  require that  $\dot{\phi}=0$. However, as it  is shown  in appendix \ref{app-kess},  when
$\dot{\phi}=0$ the dynamics of linear perturbations is pathological: both
the kinetic and mass terms vanish signalling strong
coupling; in addition, both $\delta p$ and $\delta \rho$ also vanish.
In the case $K$ is shift symmetric,
the kinetic term  for the scalar field perturbation is not identically zero, however this time the speed
of sound is zero unless higher derivative terms are
introduced~\cite{Arkani-Hamed:2003pdi}. A possibility that will be not
discussed here is to consider a scalar tensor
theory where the scalar field is not minimally coupled to gravity (see
for instance~\cite{Gubitosi:2012hu,Frusciante:2019xia,Crisostomi:2018bsp});
we just note that  passing
solar system tests~\cite{Will:2014kxa} requires some sort
  of screening
   mechanism and the measurement of the propagation speed of gravitational waves~\cite{LIGOScientific:2017zic} put significant
constraints~\cite{Lombriser:2016yzn,Ezquiaga:2017ekz,Creminelli:2017sry}
on such theories.

Before proceeding further, one might argue that focusing on a
background that saturates  the null energy condition is not terribly important phenomenologically.  After all, though our
Universe is dominated by dark energy, different subdominant components are
present and thus $w$ is not exactly one. The point is that  if
the equation of state of dark energy is $w=-1$, the total energy
density  $\bar \rho_{tot}$   and the total pressure $\bar p_{tot}$ are
such that the value of $\bar
\rho_{tot} +\bar p_{tot}$ will get closer and closer to zero as times
goes by. The only case  where the present
discussion is not phenomenologically relevant is  when
$w \neq -1$. 
In such a  case a simple scalar
field theory provides a compelling and simple  viable model of 
dynamical dark energy. 

A different avenue is to  consider dark energy with $w=-1$ as a self-gravitating medium more
general than a perfect fluid. In general, a perfect fluid (see~\cite{Kovtun:2012rj,Andersson:2020phh} for
recent reviews) can be
described in terms of three degrees of freedom (DoF) obtained from the
decomposition of the fluid velocity into a longitudinal and
transverse part. All the fluid properties can be derived from an action principle  based on three scalar fields
$\{ \Phi^a \;, \;  a=1,2,3
\}$~\cite{Matarrese:1984zw,Leutwyler:1993gf,Leutwyler:1996er} that
can be interpreted as the Eulerian coordinates of a fluid element; for a
recent discussion see~\cite{Ballesteros:2012kv,Ballesteros:2016kdx}. The
Lagrangian $U$ can be taken as a function of
\be
 b=(\text{Det}[B^{ab}])^{1/2} \, , \qquad  B^{ab}
= g^{\mu \nu} \, \de_\mu \Phi^a \, \de_\nu \Phi^b  \, , 
\qquad a,b=1,2,3 \, .
\label{bdef}
\ee
The
Lagrangian $U(b)$ has
a large internal symmetry corresponding to volume preserving internal
diffeomorphisms
\be
\Phi^a \to \Psi^a(\Phi^b),   \qquad \qquad \text{Det} \left(
  \frac{\de \Psi^a}{\de \Phi^b} \right)=1 \, ,
\label{svolprd}
\ee
and it describes a perfect barotropic fluid with 4-velocity
\be
u^\mu = -\frac{\epsilon^{\mu \nu \alpha \beta} }{6 \, b \, \sqrt{- g}}
  \epsilon_{abc} \, \de_\nu \Phi^a \,  \de_\alpha \Phi^b \,  \de_\beta
  \Phi^c \, , \qquad u^2=-1 \, ,
  \ee
 and $u^\mu \de_\mu \Phi^a
=0$. The EMT is given by (\ref{pfemt}) with
 \be
 p= U - b \, U_b \, , \qquad \rho = -U \, .
 \ee
The volume preserving reparametrization symmetry (\ref{svolprd})
implies that only the volume of fluid elements matters in a physical
configuration. A more general fluid system can
be obtained by adding a superfluid component whose velocity is the
gradient of an additional scalar field $\Phi^0$. As a
  result, two new operators with a single derivative acting on the
  scalar fields and invariant under the
volume preserving reparametrization symmetry (\ref{svolprd}) exist
\be
y= u^\mu \de_\mu \Phi^0 \, , \qquad 
\chi=(-g^{\mu\nu}\de_\mu\Phi^0\de_\nu\Phi^0)^{1/2} \, .
\ee
The velocity of the superfluid component has zero vorticity and is given by
\be
v_\mu =\chi^{-1} \, \de_\mu \Phi^0 \, .
\ee
The Lagrangian  of the form $U(b,y,\chi)$ describes a fluid-superfluid system.
The most general  non-dissipative self-gravitating medium can
be described by the same  four scalar fields by giving up the
large symmetry  (\ref{svolprd}) and requiring invariance only under
internal spatial rotations
\be
\Phi^a \to {\cal
      R}^a_b \, \Phi^b, \, \qquad  a,b=1,2,3  \qquad {\cal R} \in
    SO(3).
 \label{introt}
\ee
The additional operators that break  (\ref{svolprd})  and are
invariant under (\ref{introt}) can be chosen as
 \be
\tau_1 =  \text{Tr}(\pmb{B}) \, , \qquad     \tau_Y= \frac{\text{Tr}(\pmb{B}^2)}{\tau_1^2} \, , \qquad \tau_Z
= \frac{\text{Tr}(\pmb{B}^3)}{\tau_1^3} \, ;
\label{operator}
\ee
where $\pmb{B}$ is the 3x3 matrix with matrix elements $B^{ab}$ given
in  (\ref{bdef}).
In the framework of effective field theories~\cite{Dubovsky:2011sj,Son:2005ak,ussgf,Celoria:2017bbh},   we arrive to the action for the  most general non-dissipative
self-gravitating medium given in terms of four scalar fields
 $\{\Phi^A , \; A=0,1,2,3 \}$ of the form~\footnote{According to our
  notations $\plm^2=( 8 \pi G)^{-1}$.}
 \be
S_{DE}= \, M_{pl}^2 \, \int d^4x \, \sqrt{-g} \, U(b, y, \chi, \tau_Y,
\tau_Z) \, .
\label{ssol}
\ee
The action (\ref{ssol}) is the  leading order term in a derivative expansion and it is a sort of generalised
k-essence with the symmetries (\ref{introt}) and  $\Phi^A \to \Phi^A +
    \text{constant}$,  and it will be our model for a dynamical
    cosmological constant.
    
The energy-momentum tensor (EMT) has the form
  \be
  \plm^{-2} \, T_{\mu \nu} = (U- b \, U_b) g_{\mu \nu}  + \left(y \, U_y -b \, U_b \right) u_\mu \,
  u_\nu +\chi \, U_\chi \, v_\mu \,
  v_\nu  +Q_{\mu \nu}^{(Y)} \, U_{\tau_Y} +Q_{\mu \nu}^{(Z)} \,
  U_{\tau_Z}  \, ;
  \label{EMT}
  \ee
  with
  \bea
  && v_\mu =\chi^{-1} \, \de_\mu \Phi^0 \,  ; \\
  && Q_{\mu \nu}^{(Y)} = 2 \left( \frac{1}{\tau_1^2 }\, \de_\mu 
    \Phi^a \, \de_\nu \Phi^b \, B^{ab} - \frac{\tau_Y}{\tau_1} \de_\mu 
    \Phi^a \, \de_\nu \Phi^a  \right) \, ; \\
 && Q_{\mu \nu}^{(Z)} =3 \left( \frac{1}{\tau_1^3} \de_\mu 
    \Phi^a \, \de_\nu \Phi^b \, \left( B^2 \right) ^{ab} -
    \frac{\tau_Z}{\tau_1} \de_\mu 
    \Phi^a \, \de_\nu \Phi^a\right)  \, .
  \ea
  When $\Phi^a$ fluctuates around a background proportional to 
 $\vec{x}$, while  $\Phi^0$ has 
    a time-dependent background, the EMT describes a medium with
    mechanical and thermodynamical properties determined by the
    internal symmetries of the action (\ref{ssol}), as discussed
    in~\cite{ ussgf,Celoria:2017bbh}.
    The action (\ref{ssol}) is also related to massive gravity~\cite{ussgf}.
In flat space or in a spatially flat FRW spacetime  we have the following background values
\be
\begin{split}
  &\bar\Phi^a=x^a \, , \qquad \bar \Phi^0=\phi(t) \\
& \bar b =
  \bar \chi = \bar y=1 \qquad  \bar{u}_\mu = \bar{v}_\mu \, , \qquad 
  \bar Q_{\mu \nu}^{(Z)} =\bar Q_{\mu \nu}^{(Y)} =0 \, .
\end{split}
\label{bckval}
  \ee
Thus, the background EMT is the one of a perfect fluid with
\be
\bar \rho =-U +\bar \chi \, U_\chi  + \bar y \, U_y \, , \qquad \bar p=U -
\bar{b} \, U_b \, .
\label{prhobkg}
\ee
Depending on che choice of $U$, different equations of state for the
medium can be considered; for instance one can take  
\be
U(b,y,\chi, \tau_Y, \tau_Z) \equiv b^{1+w} \, U_{w}(b^{-w} \,  \chi ,
\;b^{-w} \, y,\;  \tau_Y, \;\tau_Z) \, ,
\label{Ude}
\ee
then from (\ref{prhobkg}) one gets that $\bar p = w \, \bar \rho$.

In general, the linear dynamical stability  in Minkowski space and in a FRW Universe is closely related to the
equation of state of the medium~\cite{Dubovsky:2005xd,Celoria:2017hfd, Celoria:2017idi}.
In flat space, exploiting internal and spatial rotational invariance,
the fluctuations  $\pi_0$ and $\pi_l$ of the scalar fields around their
  background configurations are defined as~\footnote{In Minkowski
  space one can set in (\ref{bckval}) $\phi(t)=t$.}
\be
\Phi^0=t + \pi_0 \, , \qquad \Phi^a= \delta^a_i  \left( x^i+ \de_i
  \pi_l +\pi_T^i\right) \, ,  \qquad \qquad \de_i \pi_T^i=0 . 
\ee
The dynamics of the vector modes $\pi_T^i$ will be studied in section
\ref{sect-vector}. The fields $\pi_0$ and $\pi_l$ can be interpreted as the Goldstone
boson for broken translation.

Before digging into the study of the
general case,  one may wonder whether all the four scalar fields are
mandatory, namely if dealing with the most general medium is really
needed. By turning off all operators excepts $\chi$, we
  get the Lagrangian $U(\chi)$ that describes a
perfect irrotational fluid and we are back to the case of a shift
symmetric k-essence already discussed (see
also~\cite{Ballesteros:2016kdx}). When only $b$ is present, only the
fields  $\{
\Phi^a, \; a=1,2,3\}$ are needed; unfortunately when $w=-1$ both the
longitudinal $\pi_l$ and the transverse
vector $\pi_T^i$  do not propagate~\cite{Ballesteros:2016kdx}. Consider
next the case where only the
operators $b$ and $y$ are present: $U(b,y)$ has still the large
internal symmetry (\ref{svolprd}) and it represents a non-barotropic
perfect fluid; again the dynamics of transverse and longitudinal
modes is pathological when
$p+\rho=0$~\cite{Ballesteros:2016kdx}. Finally, let us consider the
case of pure solid-like medium; the internal symmetry (\ref{svolprd})
is not present, the Lagrangian is of the  form
$U(b,\tau_Y, \tau_Z)$ and only the three scalar fields $\{
\Phi^a, \; a=1,2,3\}$ are needed.  
From the expansion at the quadratic level of (\ref{ssol})  in which $y$ and $\chi$ are omitted we get~\footnote{The dot denotes the time derivative, fields are
  expressed in the Fourier basis and depend on time and on the  spatial momentum $\vec{\pmb{k}}$; thanks to
  the rotational invariance only $|\vec{\pmb{k}}|=k$ is present in the
  Lagrangian. To simplify the notation we use same name
  $\varphi(t,\vec{x})$ for the field and its spatial Fourier transform  $\hat
  \varphi(t,\vec{k})$ defined by
  $$
  \varphi(t,\vec{x}) =\int \frac{d^3k}{\left(2 \pi \right)^3} \, e^{i
    \vec{k} \cdot \vec{x} } \; \hat   \varphi(t,\vec{k}) 
 $$}  
\be
{\cal L}^{(2)}_{\text{solid}} = \frac{\left(\bar p+\bar \rho
  \right)k^2}{2} \dot{\pi_l}{}^2 +k^4 \, \plm^2 \, \left(M_4-M_2 \right)
\pi_l^2 \, ;
\ee
the parameters $\{M_A, \; A=0,1,2,4\}$ can be expressed in terms of
the derivatives of $U$ whose form can be found in appendix
\ref{lan}. Once again, when $\bar p+\bar \rho=0$ , the kinetic term
vanishes. As a result, in order to have a non-pathological dynamics for the  dark energy sector
perturbations,  all four scalar fields are
needed and the action (\ref{ssol}) describes  a medium that is not a perfect fluid nor a
perfect solid but a combination of a solid with a superfluid component (supersolid). 

Let us now consider the general case and focus on scalar modes; their dynamics at the
linear level  is described by the following Lagrangian  in Fourier space obtained from the
quadratic  expansion of (\ref{ssol})
\be
{\cal L}^{(2)}= \frac{\plm^2}{2} \left[ \dot{\varphi}{}^t {\cal K} \dot{\varphi}+ 2 \,
  \varphi^t {\cal
  D} \dot{\varphi} - \varphi^t {\cal M} \varphi\right] \, ,  \qquad \qquad \varphi^t=(k^2 \, \pi_l , \, k \,
\pi_0)  \, ;
\label{lang}
\ee
where 
\be
\begin{split}
& {\cal K}=\ \left(
\begin{array}{cc}
 \frac{M_1+ \plm^{-2} \left(\bar p+\bar \rho \right) }{k^2} & 0 \\
 0 & \frac{2\, M_0}{k^2} \\
\end{array}
\right), \, \qquad {\cal D} =\left(
\begin{array}{cc}
  0 & \frac{\left(M_1-2 M_0\right)}{2 k} \\
 -\frac{\left(M_1-2 M_0\right)}{2 k} & 0 \\
\end{array}
\right), \\[.2cm]
& {\cal M} = \left(
\begin{array}{cc}
 \frac{2}{3}  \left(2 M_2-3 M_0\right) & 0 \\
 0 & - M_1 \\
\end{array}
\right) \, .
\end{split}
\label{matrices}
\ee
 When $w > -1$, e.g. $\bar \rho+ \bar p >0$, stability is rather
standard and the Hamiltonian is positive definite~\cite{Celoria:2017hfd}. However this is not the case when
$w=-1$; the condition for ${\cal K} >0$ conflicts with ${\cal M}>0$.
The best one can do is to reduce the dynamics to independent ``normal modes'' of the form $\exp \left({c_{s 1/2}
  t}\right)$ and  require that the sound speeds $c_{s1/2}$ are real,
avoiding instabilities. As discussed in detail in~\cite{Comelli:2022evf}, the procedure is the following:
 by a suitable field redefinition, one can always put the Lagrangian
(\ref{lang}) in the standard form in which ${\cal K}$ and ${\cal M}$
are diagonal
\be
{\cal D} \to D=\begin{pmatrix} 0 & d \\- d & 0 \end{pmatrix}  \, ,
\qquad {\cal M} \to M=\begin{pmatrix} m_1^2 & 0 \\ 0 &
    m_2^2 \end{pmatrix} \, ;
  \ee
  with
  \be
  d = \frac{k \left(M_1-2 M_0\right)}{2 \sqrt{2} \sqrt{M_0}
   \sqrt{M_1}} \, , \qquad m_1^2 =\frac{2 k^2 \left(2 M_2-3 M_0\right)}{3
   M_1} \, , \qquad m_2^2=-\frac{k^2 M_1}{2 M_0} \, .
 \label{canpar}
  \ee
The system (\ref{lang}), studied in~\cite{Comelli:2022evf},  is rather
peculiar due to the presence of the antisymmetric matrix ${\cal D}$ that mixes $\varphi$ with its time
derivative and falls under the class of gyroscopic
systems~\cite{gyro}. By using a  suitable canonical
transformation $( \Pi, \varphi) \to ( \Pi_c, \varphi_c)$, the Hamiltonian
$H$ associated to (\ref{lang}) can be diagonalized, however
 its form
crucially depends on the signs of $m_1^2$ and $m_2^2$.  The standard
case of a positive definite energy is realised when  $m_{1/2}^2>0$
and the diagonal form of $H$ is the sum of two harmonic
oscillators. Unfortunately, this is impossible when $w=-1$; indeed,
taking ${\cal K}>0$ which is equivalent to  $M_1, \; M_0>0$, leads to 
$m_{1/2}^2<0$. The Hamiltonian can be written
as the {\it difference} of two harmonic oscillators~\footnote{Actually the ``standard'' form is
  obtained with a further canonical transformation $\Pi_{ca}=
  \omega^{-1/2}_a \, P_a$, $\varphi_a=  \omega^{1/2}_a \, q_a$  with $a=1,2$.}
\be
H= \frac{\omega_1}{2} \left( \Pi_{c1}^2 +\varphi_{c1}^2\right) -
\frac{\omega_2}{2} \left( \Pi_{c2}^2 +\varphi_{c2}^2\right) \, ,
\label{hdiag}
\ee
where
\be
\omega^2_{1,2} = \ha \left (4 d^2+m_1^2+m_2^2 \pm \sqrt{\left( m_1^2
      +m_2^2 + 4 d^2\right)^2 - 4 m_1^2 m_2^2}\right) \, .
\ee
When the Hamiltonian is not positive definite,  linear
stability requires that~\footnote{In~\cite{Comelli:2022evf}  it was called anomalous
  stability.}  $\omega_{1/2} >0$ and  leads to 
\be
m_{1,2}^2 < 0, \, \qquad d^2 \geq \frac{\left( \sqrt{-m_1^2} +
    \sqrt{-m_2^2} \right)^2}{4} \, ;
\label{anstab}
\ee
By considering (\ref{canpar}), from (\ref{anstab}) one gets
\be
M_0> \frac{2}{3} M_2 \, , \qquad  M_1>0 , \qquad  M_1+\sqrt{M_0^2-\frac{2 M_0
    M_2}{3}}<M_0 \, .
\label{stab}
\ee
When $w=-1$, necessary conditions for stability are:
\begin{itemize}
\item
 the dynamical cosmological constant must have a non-trivial anisotropic
 stress~\footnote{The sources of anisotropic stress are $Q^{(Y)}_{\mu \nu}$ and $Q^{(Y)}_{\mu
     \nu}$ and the first non-vanishing contribution starts with $\de_i
   \pi_j + \de_j \pi_i$ and their contribution to the EMT is
   proportional to $U_{\tau Y}$ and $U_{\tau z}$ or equivalently to
   $M_2$, see eq. (\ref{masses}).} 
 $M_2 \neq 0$;
\item
 the total Hamiltonian cannot be positive definite.
\end{itemize}
An alternative equivalent form of the above inequalities is obtained
by setting~\footnote{We  order conventionally $c_{s1}$ and $c_{s2}$ such that $c_{s2} < c_{s1}$.}
\be
\omega_1^2 = k^2 \, c_{s1}^2  \, , \qquad \omega_2^2 = k^2 \, c_{s2}^2
\, ;
\ee
then
\be
0< c_{s1}^2 \; ,  c_{s2}^2 <1 \, , \qquad M_2 >0 \, , \qquad M_1 >0 \, .
\label{staba}
\ee
Once (\ref{stab}) or (\ref{staba}) are satisfied,  the solutions of the equations of
motion show the standard oscillator-like behaviour for both $\pi_0$ and $\pi_l$.
Let us point out  that  taking  the limit of zero anisotropic stress, namely
$M_2 \to 0$, one gets
\be
\lim_{M_2 \to 0}  \omega_{1/2}^2= - k^2 \, ,
\ee
which leads  to an exponential instability. Such a limit is naturally
obtained by taking the Lagrangian for the medium of the form
$U(b,y,\chi)$; the symmetry  (\ref{svolprd}) associated with  a
perfect fluid is present and the Lagrangian describes a coupled
system of a fluid and a superfluid. It follows that the solid
component is essential to avoid exponential instability. The results are summarised in table \ref{tablemed}.
%%%%%
\vskip .5cm
\begin{table}
  \caption{The dynamical properties of fluctuation of the various media
    considered when the background pressure and density satisfy
    $\bar p +\bar \rho=0$. The number of degrees of freedom is split
    into scalar and transverse vector modes.}
  \begin{tabular}{|p{3cm}|p{3.5cm}|p{1cm}|p{3cm} ||}
\hline
    \hline
    Lagrangian &  Medium Type  &  DoF  &  Properties\\
    \hline    \hline
    $ U(\chi)$ & superfluid & 1 &  zero kinetic term\\
    \hline
   $  U(b)$ & perfect barotropic fluid  & 2+1 &  zero kinetic term\\
    \hline
    $  U(b,y)$ & perfect fluid  & 2+1 &  zero kinetic term\\
    \hline
    $  U(b,y,\chi)$ & fluid/superfluid  & 2+2 &  zero kinetic term\\
    \hline
    $  U(b, \tau_Y,\tau_Z)$ & solid  & 2+1 &  zero kinetic term\\
    \hline
    $  U(b, y, \chi, \tau_Y,\tau_Z)$ & supersolid  & 2+2 &  healthy dynamics\\
     \hline    \hline
  \end{tabular}
  \label{tablemed}
  \end{table}
%\end{center}
%\vskip .5cm
%%%%
It is worth to point out that  the Hamiltonian (\ref{hdiag}) is closely
related to the one of  the Pais-Uhlenbeck
oscillator~\cite{Pais:1950za}. Originally, Pais and Uhlenbeck studied
a higher derivative system as a model, trying  to improve the high energy
behaviour of interacting relativistic quantum field theories. The fate
of system like (\ref{hdiag}) at the classical and quantum level when
interactions are introduced has become a subject of a number of recent
studies. In the present investigation we are interested to the  {\it classical} behaviour of a
 system with an Hamiltonian of the form
  (\ref{hdiag}). The
question is what happens to the perfectly stable system like
(\ref{lang}) when it is coupled with other degrees of freedom. The
fear is that exponentially fast instabilities can
  develop by turning on a small interaction that allows energy  exchange with 
a system that has  an unbounded from bellow Hamiltonian; however such instabilities are not necessarily
present~\cite{Smilga:2017arl,Gross:2020tph,Deffayet:2021nnt}. 
In the contest of dark energy, gravity naturally provides an
indirect interaction between dark energy and standard matter. Actually
gravitational (Jeans) instability triggers structures formation; in particular such mechanism is very
efficient during matter domination while it stops when the Universe
enters in a phase of cosmological constant domination. The natural
question is what happens when the cosmological
constant is replaced by the dynamical model of dark energy described by
(\ref{ssol}) with $w=-1$. The rest
of the paper is devoted to answer this question by using linear
cosmological perturbations to study the impact of DCC on structure
formation and on the propagation of gravitational waves and vector
modes.

\section{Cosmological Perturbations: Dark Energy Domination}
\label{cosmpert}
Consider now the evolution of cosmological perturbations in a Universe
dominated by the dark energy component described by (\ref{ssol}) with
$w=-1$; standard matter and radiation will give a subdominant
contribution that  will be neglected here. 
By using the Newtonian gauge and the conformal time $t_c$ for the
scale factor $a$, the scalar part of the
metric perturbations can be written as 
\be
ds^2 = a^2 \left[ -(1+ 2 \, \Psi) dt_c^2+  (1-2 \, \Phi)
  \delta_{ij} \, dx^i dx^j \right] \, ,
\label{gpertnewt}
\ee
while the scalar perturbations of the dark energy sector read
\be
\Phi^i= x^i + \de_i \pi_l \, , \qquad \Phi^0= \phi(t_c) + \pi_0 \, .
\label{backval}
\ee
In the scalar sector there are two independent modes that can be taken
to be $\pi_l$ and $\pi_0$.
When  $w=-1$ is set in (\ref{Ude}), the parameters $\{ M_A \, , A=0,1,2,3,4
\}$ are time independent and moreover they satisfy the following relations~\footnote{This can be seen as the consequence of a  Lifshitz
  scaling symmetry~\cite{Endlich:2012pz,Celoria:2017idi}.}
\be
\frac{M_4}{M_0} =1 \, , \qquad M_2 -3 (M_3-M_4) =0 \, .
\label{Mrel}
\ee
The Einstein equations are given by
\be
G_{\mu \nu} = 8 \, \pi \, G \, T_{\mu \nu} \; ,
\ee
where the RHS is given by (\ref{EMT}). At the  linear level, the EMT can be written as
\be
T_{\mu \nu}^{(DE)} = \bar T_{\mu \nu}^{(DE)}  +T_{\mu
  \nu}^{1(DE)} \, ;
\ee
where $\bar T_{\mu \nu}^{(DE)} $ is the EMT of perfect fluid with
background pressure and energy density given by (\ref{prhobkg}); for a dark
energy dominated era $T_{\mu \nu} \approx T_{\mu
  \nu}^{(\text{DE})}$. The explicit form of the linear order
perturbation $T_{\mu \nu}^{1(DE)}$  of $T_{\mu \nu}^{(DE)} $ can be
found in appendix~\ref{app-EMT}. We stress again that, when perturbations
are taken into account, $T_{\mu \nu}^{(DE)} $  has not the form of a perturbed perfect fluid. The presence of the operator $\chi$
built out $\Phi^0$ breaks the internal symmetry: $\Phi^0 \to \Phi^0 + f(\Phi^a)$;
when this is the case $M_1\neq0$ and then the  two scalar
perturbations $\pi_l$ and $\pi_0$ both propagate. Notice that, when $M_1
\neq 0$, even if $w=-1$,  still $T_{0i}^{1(DE)}\neq 0$ and the medium has a
non-trivial velocity.  Moreover $T_{\mu \nu}^{(DE)} $ features a non-trivial anisotropic stress proportional
to $M_2$. With respect to the previous section the change
  in the equations of motion for $\pi_l$ and $\pi_0$ are due to the
  effect of the gravitational background and the presence of the
  gravitational fluctuations. At the background level we get
the standard relations
\be
3 \, {\cal H}^2 = 8 \, \pi \, G a^2 \, \bar{\rho} \, , \qquad {\cal
  H}^2 +2 \, {\cal H}' =  - 8 \, \pi \, G a^2 \, \bar p \,  \, ;
\ee
where ' denotes the derivative with respect to the conformal time $t_c$ and ${\cal
  H} =a'/a$ is the Hubble parameter in conformal time. In accordance
with (\ref{prhobkg}), $\bar \rho$ and $\bar p$ are
given by
\be
\bar \rho= -\frac{a U-\phi ' \left(U_y+U_{\chi } \phi '\right)}{8 \pi
  a G} \, , \qquad \qquad  \bar p = \frac{a^3 U-U_b}{8 \pi  a^3 G}\, .
\ee
The conservation of the EMT at the
background level gives
\be
\phi ''-\frac{\left(M_0+3 M_4\right) \mathcal{H} \phi
  '}{M_0}=0 \, .
\label{phieq}
\ee
In the case of $w=-1$, $ {\cal
  H}^2 - \, {\cal H}'=0$ and the relations (\ref{Mrel}) hold; from
(\ref{phieq}) we get
\be
\phi'=a^4 + \text{constant} \, .
\ee
From now on we will focus on the case $w=-1$ and thus the  EMT is
the dynamical generalisation  of a cosmological constant. By using the linearised Einstein equations, one can express the metric perturbations in terms of
$\pi_l$ and $\pi_0$; in particular
\be
2 \, a^2 \, M_2 \, \pi _l -\Phi+\Psi  =0 \, ,
\label{anieq}
\ee
and
\be
\Phi= \left( 2 a^2 M_0+k^2 \right)^{-1}\left[a^2 \pi _l M_0 \left(k^2-2 a^2 M_2\right)-\frac{3}{2} a^2 M_1
   \mathcal{H} \pi _l'+\frac{3 \pi _0 M_1 \mathcal{H}}{2 a^2}-\frac{M_0
   \pi _0'}{a^2} \right] \, .
\label{Phieq}
\ee
As expected, the presence of the solid component triggers a
non-vanishing anisotropic stress even in the scalar sector; as a
result, the difference between the two scalar Bardeen potentials is
proportional to $\pi_l$, that is relevant for the propagation of CMB photons. 
The EMT conservation, together with (\ref{anieq}) and (\ref{Phieq}), can be used to get the following dynamical
equations for $\pi_l$ and $\pi_0$
\bea
&&\pi _l'' + 2 \mathcal{H} \left(2-\frac{3 a^2 M_0}{2 a^2
   M_0+k^2}\right) \pi _l'  +2 \frac{ \left[10
   a^2  M_0 M_2+(2  M_2 -3 M_0) k^2\right]}{3 M_1 (2 a^2
 M_0 + k^2)} k^2 \, \pi _l +\frac{6 M_0 \mathcal{H}}{a^2
 \left(2 a^2 M_0+k^2\right)}  \pi _0 \nb \\
&&+\frac{ \left[k^2 (2  M_0- M_1) -2 a^2 M_0
   M_1\right]}{a^4 \, M_1 (2 a^2 M_0 + k^2)} \pi _0'=0
\, ; \label{eqpil}\\[.2cm]
&&\pi _0'' - \left(\frac{k^2}{2
   a^2 M_0+k^2}+3\right) \mathcal{H} \, \pi _0'-  \left[a^2 M_1 \left(\frac{3 \mathcal{H}^2}{2
   a^2 M_0+k^2}+1\right)+\frac{k^2 M_1}{2
 M_0}\right] \pi _0\nb \\
&&\frac{2 a^6 \mathcal{H} \left[10 a^2 M_0 M_2+k^2 \left(4
      M_2-M_0\right)\right]}{2 a^2 M_0+k^2} \, \pi_l \nb \\
&&+
   \left[\frac{a^4 k^2 \left(M_1-2 M_0\right)}{2
   M_0}+a^6 \left(\frac{3 M_1 \mathcal{H}^2}{2 a^2
   M_0+k^2}+M_1+2 M_2\right)\right] \pi _l' =0 \, . \label{eqpi0}
\ea
The presence of terms with the momentum $k$ in the
  denominators is due to the eliminations of $\Psi$ and $\Phi$ in
  favor of $\pi_l$ and $\pi_0$ using (\ref{anieq}) and (\ref{Phieq}). The
scale factor during dark energy domination, expressed in conformal
time $t_c$, is the one of de Sitter spacetime
\be
a(t_c) = \frac{1}{1-H_0 t_c} \, , \qquad t_c \in [0, \, H_0^{-1}) \, . 
\ee
The  present epoch corresponds to an epoch of dark energy domination
that starts at the conventional time $t_c=t_{c0}=0$ and follows matter domination~\footnote{We neglect the
  sub-leading effect of photons and baryons.}. The value of the
constant $H_0$ is set
to be the present value of the Hubble parameter. To simplify the form of the scale factor is convenient
to redefine the conformal time according to: $t_c= \tau +H_0^{-1}$ with $\tau
\in [-H_0^{-1}, \, 0)$; then
\be
a(\tau) = -\frac{1}{H_0 \, \tau} \, .
\label{ads}
\ee
The form (\ref{ads}) for the scale factor in dS corresponds to
a spatially flat section; this is the most natural choice, given  the
overwhelming evidence that the spatial
curvature is negligible before dark energy domination. 

At level of linear cosmological perturbations, the effect
  of gravity manifests itself as non-local modifications (in space)  of the
  equations of motion for $\pi_l$ and $\pi_0$ with respect to  the ones
founded in the previous section. As expected, in the very small scale
  limit (large $k$) we recover the flat space case. % The system of equations (\ref{eqpil}-\ref{eqpi0}) can be
% decoupled by writing two independent forth-order differential equations for $\pi_0$
% and $\pi_l$ that can  be solved analytically at large
% scales.
We refer to a large scale when the physical wavelength $\lambda_{ph}\sim a/k$ is much bigger 
then the dS curvature scale $H_0^{-1}$, namely $x = k \, |\tau| \ll1$,
and  to a small scale
when $x  \gg 1$. A mode $k$ crosses
the  dS ``horizon'' when   $k |\tau| = 1$. Basically all the modes
of physical interest will cross the dS horizon, eventually. As a
reference, a comoving scale with $k=k_{eq} =\ha  10^2 H_0\sim 10^{-2}\;
\text{Mpc}^{-1}$ that has
crossed the FLRW horizon~\footnote{it should stressed that  when
  $\rho + 3 \, p \geq 0 $ the particle horizon and the Hubble horizon
  $H^{-1}$ are numerically equivalent.}  at  matter and
radiation equality, will become superhorizon again during the dS phase
at $\tau_{eq}=-2 \, H_0^{-1} \, 10^{-2}$.

It is useful to define
\be
M_A = H_0^2 \, c_A  \, , \qquad A=0,1,2,3,4  \, .
\label{cpar}
\ee

\subsection{Large Scales}
One can decouple (\ref{eqpil}-\ref{eqpi0}) by transforming them in two
fourth-order independent equations for $\pi_0$ and $\pi_l$. For large
scales, the fourth-order independent equation for $\pi_l$ is Euler-like
and reads
\be
\pi _l{}^{(4)} -\frac{4 }{t} \pi _l{}^{(3)}+\frac{2
   \left(c_2+4\right) }{t^2} \pi _l''-\frac{4 \left(3 c_2+2\right)
 }{t^3}\pi _l'+\frac{20 c_2}{t^4} \pi _l= 0 \,  
\ee
and can be easily solved
\be
\pi_l = \alpha_1 \, (-H_0 \, \tau)^{\frac{1}{2} \left(3-\sqrt{9-8
      c_2}\right)} + \alpha_2 \, (-H_0 \, \tau)^{\frac{1}{2} \left(3+\sqrt{9-8
      c_2}\right)}+ \alpha_3 (H_0 \, \tau)^2 -  \alpha_4 (H_0 \, \tau)^5
\, .
\ee
For typical values of $c_2$, the only growing mode is the one proportional
to the integration constant $\alpha_1$. For large scales, the
equation satisfied by $\pi_0$ is more complicated and takes the form
\be
\pi _0{}^{(4)}+\frac{16} {t}   \pi _0{}^{(3)} +\frac{2 \left(2 c_2+34\right)}{t^2}  \pi
   _0'' +\frac{8
   \left(c_2+9\right)}{t^3}  \pi _0'-\frac{2 \left(4 c_2^2+7 c_1 c_2+42 c_1\right) k^2}{3 c_1 t^2} \pi _0=0 \, .
 \ee
The solution of the above equation can be given in terms of
generalised hypergeometric functions; in the limit $x \ll 1$, omitting the
non-growing terms, it takes the form
\be
\pi_0 =\frac{\gamma_1}{(k \tau)^3} + \frac{\gamma_3}{(- k \tau)^{\ha
    \left(7+\sqrt{9 - 8 c_2} \right)}}  + \frac{\gamma_4}{(- k \tau)^{\ha
      \left(7-\sqrt{9 - 8 c_2} \right)}} \, .
  \ee
From (\ref{Phieq}), given the growing character of the scalar field perturbations, we also
get a growing mode for $\Phi$
\be
\Phi = \frac{H_0^2\, C_\Phi}{k^2} \, a^{\frac{1}{2} \left(\sqrt{9-8
      c_2}+1\right)} \, ;
\ee
where $C_\Phi$ is a combination of $k-$dependent integration constants 
and the  $\{ c_A \}$ defined in (\ref{cpar}); as usual, non-growing terms
have been neglected. The coupling with gravity induces a growth of the scalar perturbation
at large scales during the dS phase dominated by the DCC  (\ref{ssol}). The above result is very
different form LCDM where  $\Phi$ is
constant or decreasing during matter (radiation) domination and
$\Lambda $ domination. Superhorizon modes suffer from  gauge ambiguities, and, as discussed in
the appendix \ref{app-gaugeinv}, the fields of $\pi_l$, $\pi_0$,
$\Psi$ and $\Phi$ in the Newtonian gauge can be extended to gauge invariant quantities in a
generic gauge and thus are physical.

\subsection{Small scales}
In the opposite limit: $x \gg 1$, at the leading order, both the forth-order equations
  assume the very same form as in flat space which leads to pure oscillating solutions
  \be
  \pi_{a} = \beta_a^{(1)} \, e^{i c_{s1} \, \tau} + \beta_a^{(2)} \,
  e^{i c_{s2} \, \tau} +\beta_a^{(3)} \, e^{-i c_{s1} \, \tau} +
  \beta_a^{(4)} \, e^{-i c_{s2} \, \tau} \, , \qquad a=0, \, l \; .
\label{smallpi}
  \ee
  Of course of among the eight integration constants $\beta_a^{(i)}$, only four are
  independent and can be determined from the initial conditions for
  $\pi_0$ and $\pi_l$. As expected, we recover the oscillating behaviour of flat
  space for wavelengths much smaller than
  the dS curvature scale.
  % %
  \subsection{Numerics}

\begin{figure}[!h]
\hskip -0.1truecm    
    \includegraphics[width=.4\textwidth]{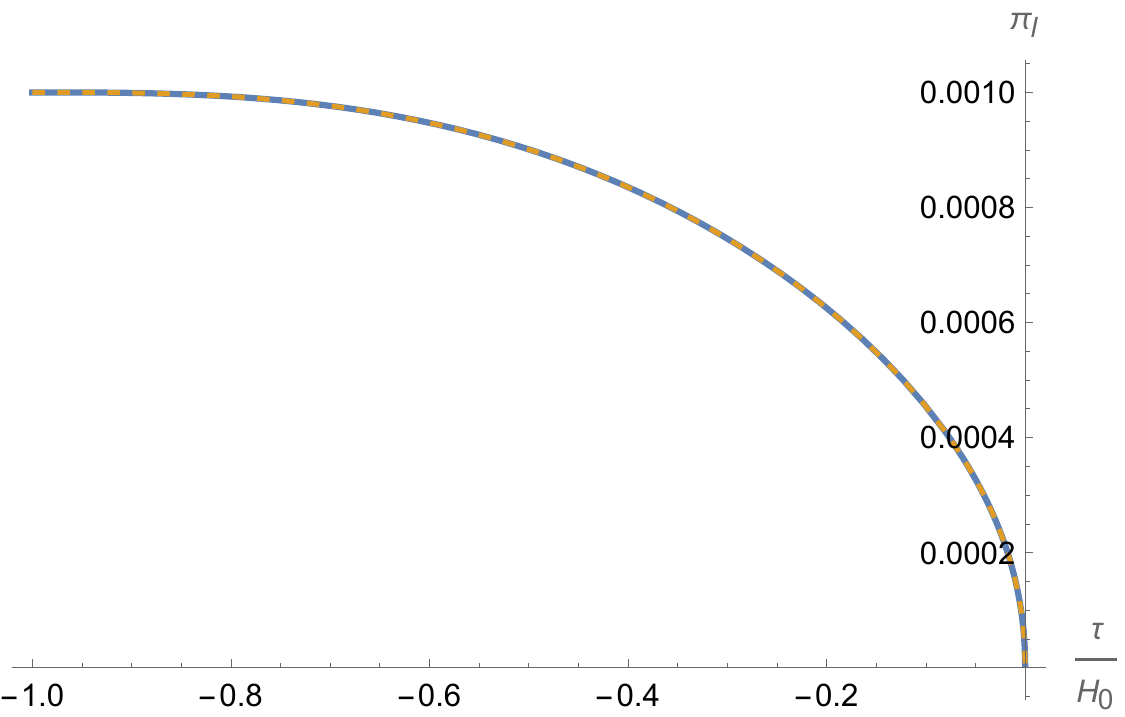}
    \includegraphics[width=.4\textwidth]{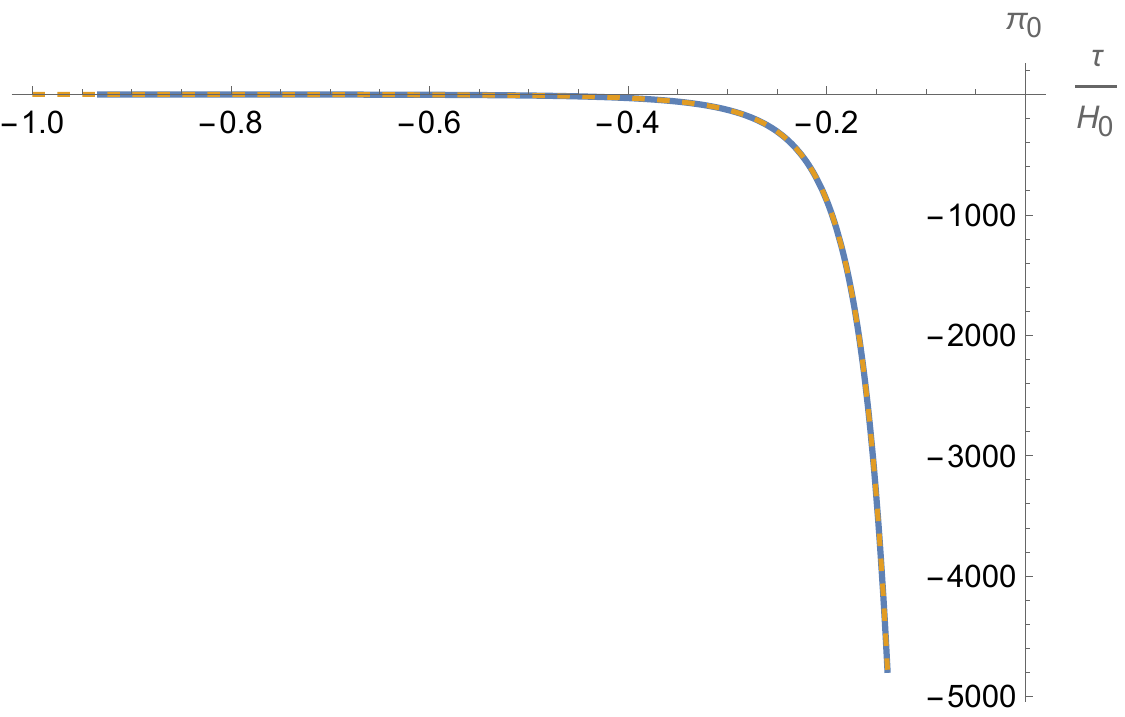}
 \caption{\label{fig-pi1} The functions $\pi_l$ and $\pi_0$ computed
   numerically (dashed) and analytically (thick) for $k=10^{-3} \, H_0$.}
\end{figure}

\begin{figure}[!h]
\hskip -0.1truecm    
    \includegraphics[width=.4\textwidth]{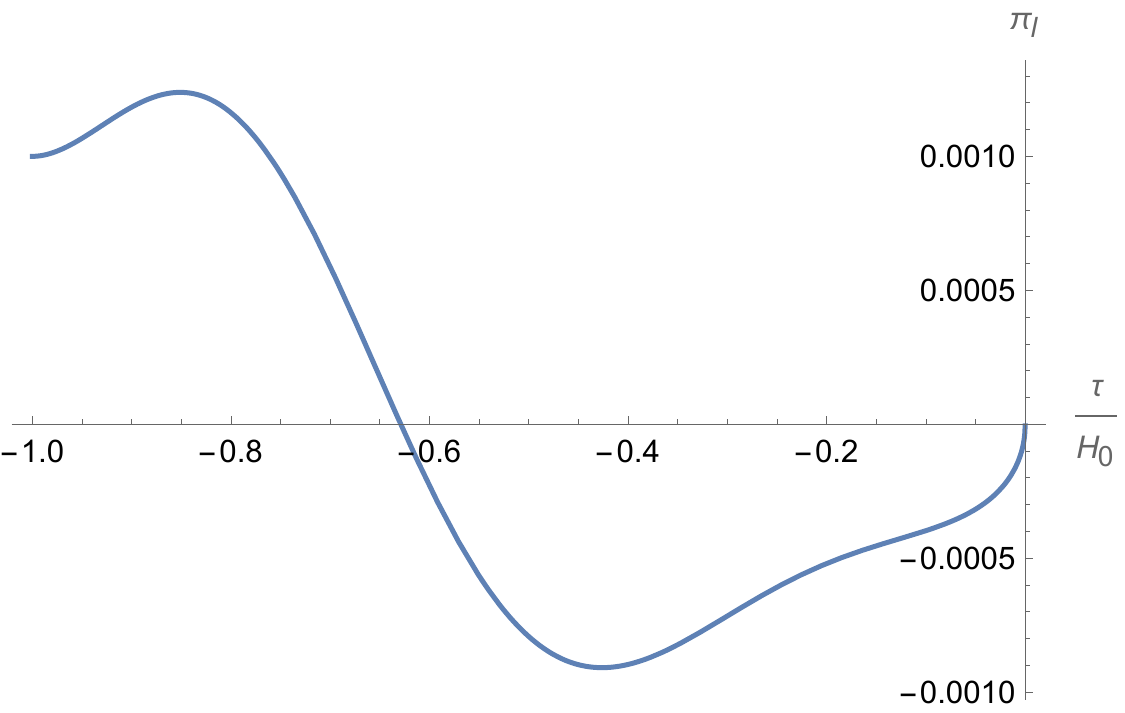}
    \includegraphics[width=.4\textwidth]{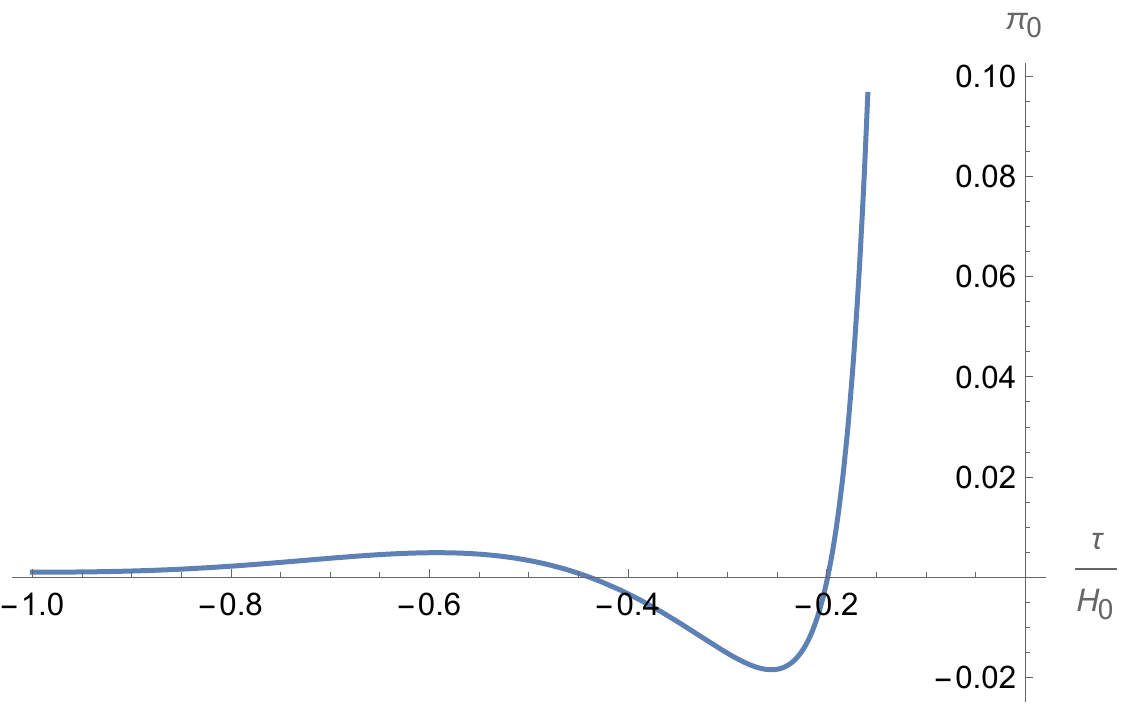}
 \caption{\label{fig-pi2} The functions $\pi_l$ and $\pi_0$ computed numerically for $k=10 \, H_0$.}
\end{figure}

\begin{figure}[!h]
\hskip -0.1truecm    
\includegraphics[width=.4\textwidth]{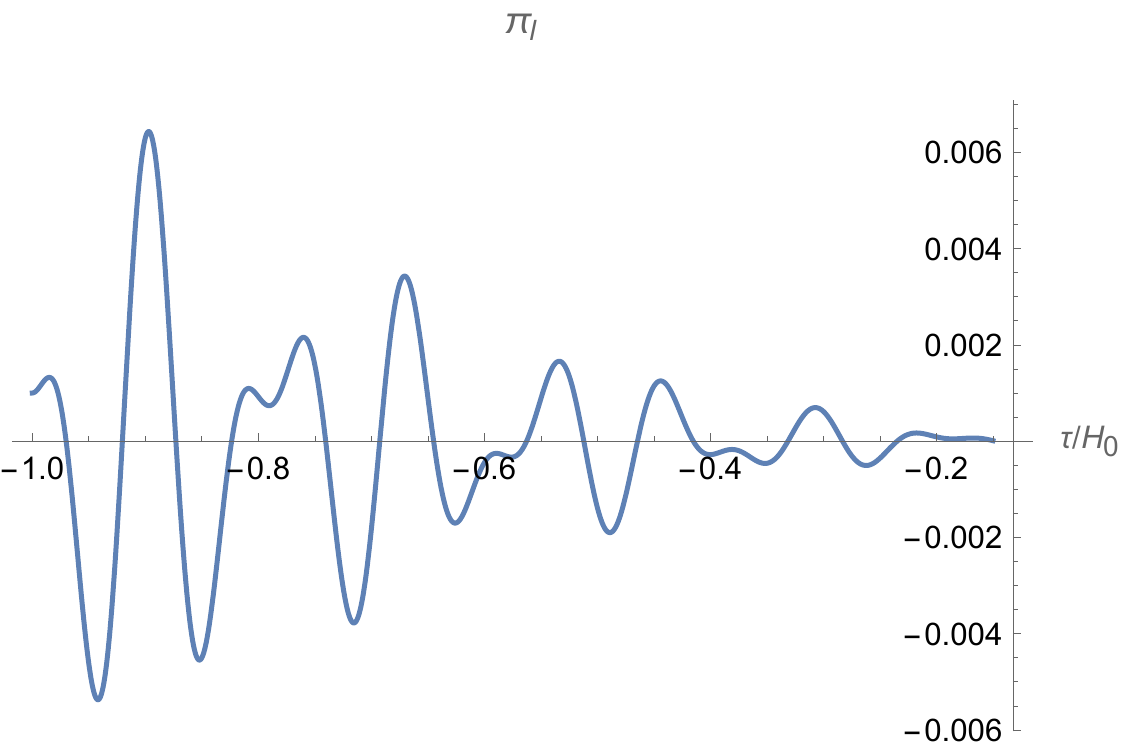}
\includegraphics[width=.4\textwidth]{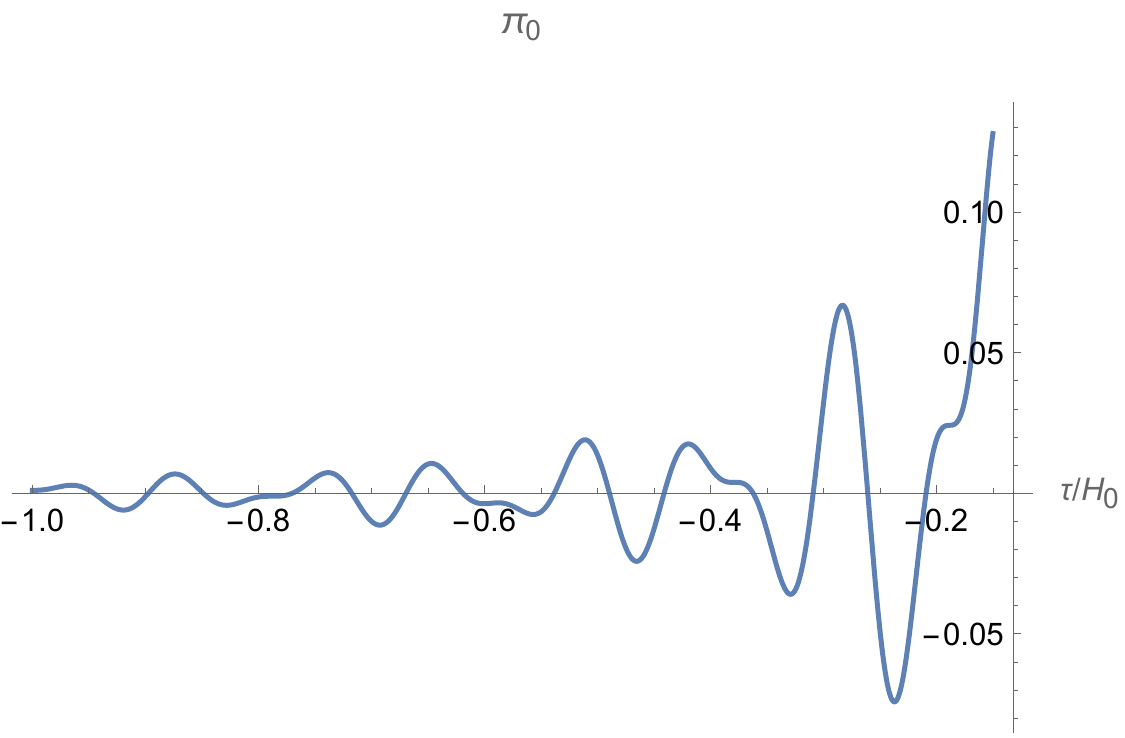}
 \caption{The functions $\pi_l$ and $\pi_0$ computed numerically for $k=10^2 \, H_0$.}
\label{fig-pi3}
\end{figure}

In order to follow the evolution of perturbations at a generic scale,
one can numerically integrate the equations of motion and compare them,  when
it is possible, with the corresponding analytical expressions. We
use  the following initial conditions:
\be
\pi_0(-H_0^{-1})=\pi_l(-H_0^{-1})=10^{-3}, \qquad
\pi_0(-H_0^{-1})'=\pi_l(-H_0^{-1})'=0 \, ,
\label{initcon}
\ee
and the numerical values of the parameters:
\be
c_0 = 0.506, \qquad c_1 = 0.266, \qquad c_2 = 0.6  \qquad \Rightarrow
c_{1s}^2=0.7142,  \quad c_{2s}^2=0.2933\, .
\label{numval}
\ee
Figure \ref{fig-pi1} shows the scalar fields perturbations for
 a large scale mode, while figure \ref{fig-pi2} shows the
case of an intermediate scale.
\\Finally,  figure \ref{fig-pi3} depicts the scalar field
perturbations for a mode of the order of $k_{eq}$.
\\Moving from  large to small scales,  consistently with the analytical solution (\ref{smallpi}), an oscillatory regime sets in. As
we shall see, the small scale oscillation of dark energy leaves an
imprint on matter perturbations. 
One might have feared that any external coupling of a system with free
Hamiltonian (\ref{hdiag}) might trigger
a catastrophic
instability; however this is not the case, at least when the system is minimally
coupled  to gravity: in an expanding universe is present only a power-law growth that resembles a
Jeans-like instability. This behavior differs 
from the standard gravitational instability in the presence of
ordinary matter, where the growth of perturbations takes place on
subhorizon scales.

\section{Structure Formation and Dark Energy}
\label{growth}
Once dark energy perturbations are known, the behaviour of standard matter perturbations with a constant equation of
state $w_m$ is found from the separate conservation of the matter's EMT, which gives
\bea
&& \delta_m '=(w+1) \left(3 \, \Phi'+  \, k^2 \, 
    v_m \right) \, ; \\
&&\left(w_m+1\right) \left[ \Psi+(1-3 w_m) {\cal H} \, v_m +v_m' \right]+\delta_m  \, w_m=0 \, ;
\ea
where $\delta_m = \delta \rho_m/\bar \rho_m $ is the matter contrast
and $v_m$  in the matter longitudinal velocity field  defined by 
\be
u^\mu_{(m)}=\bar u^\mu_{(m)}+ u^{(1)}{}^\mu_{(m)}  \, , \qquad \bar
u^\mu_{(m)}=(a^{-1},0) \, , \quad u^{(1)}{}^\mu_{(m)} =(- \Psi \,
a^{-1}, \de_i v_m + v_{(Tm)}^i) \, \, \quad  \de_i  v_{(Tm)}^i =0 \, .
\ee
For structure formation in
non-relativistic matter, one can set $w_m=0$, then
\be
\delta_m''+\mathcal{H} \, \delta_m'+k^2 \, \Psi-3 \, \Phi''-3 \,
\mathcal{H} \, \Phi' =0 \, .
\label{eqdelta}
\ee
By using the equations (\ref{anieq}-\ref{Phieq}) to express 
$\Phi$ and $\Psi$ in term of the scalar fields one gets an inhomogeneous
equation for the matter contrast with a general solution of the form
\be
\delta_m (t) =\delta_0 + \delta_{-2} \, a^{-2}+ \delta_m^{(p)}(t)  \, .
\ee
At large scales ($x \ll 1$), the particular solution $\delta_m^{(p)}$ can be obtained analytically by
using the Green method and  the
analytic expressions for $\pi_0$ and $\pi_l$;  omitting, as
usual, the decreasing and constant modes, one gets
\be
\delta_m^{(p)}= \left( \frac{H_0}{k}\right)^{2-\ha(9 - 8 c_2)^{1/2}} \, \frac{\beta_1}{(- k \, \tau)^{\ha(1+\sqrt{9 - 8 c_2})}}  +
  \beta_2 \, \left(\frac{H_0}{k}\right)^4 \, \log(a) \, ,
  \ee
with $\beta_{1/2}$ suitable constants that depend on the parameters of
the medium. The exact form for the matter
contrast valid for all scales can be obtained by solving numerically  equations
(\ref{eqpil}-\ref{eqpi0}) and (\ref{eqdelta}), taking the initial
conditions (\ref{initcon}), $\delta_m(-H_0)=10^{-3}$ and the values (\ref{numval}) for the parameters. 
\begin{figure}[!h]
  \centering
  \includegraphics[width=.45\textwidth]{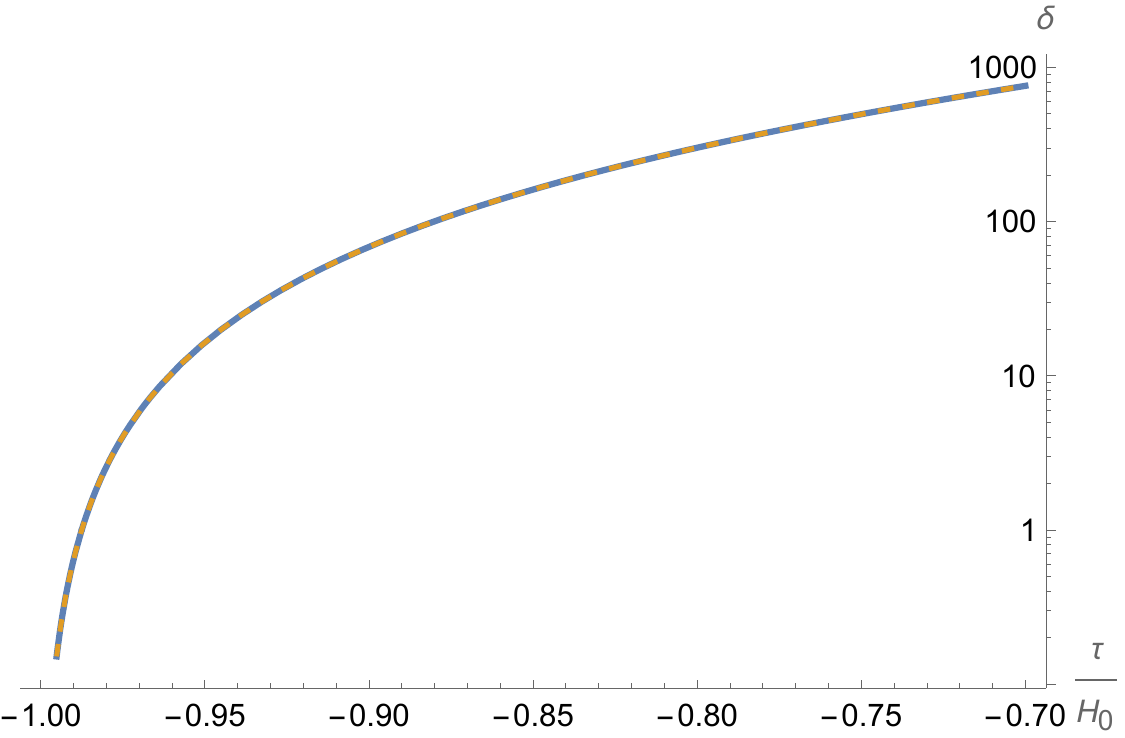}
    \includegraphics[width=.45\textwidth]{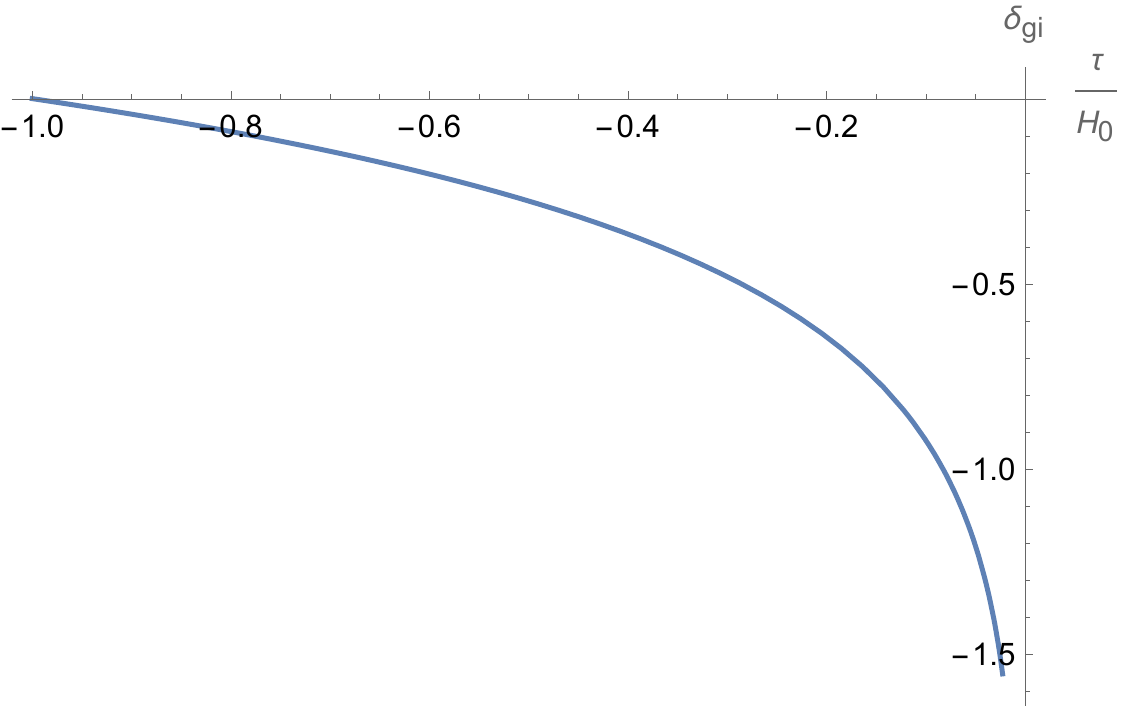}
    \caption{Matter contrast for the case $k=10^{-3} \, H_0$. On the left hand
side the dashed curve represents the numerical solution while the  thick one the
analytical form for  $\delta_m$ (log plot). On the right hand side it 
is shown the corresponding gauge invariant matter contrast $\delta_{gi}$.}
\label{fig-delta}
\end{figure}
\begin{figure}[!h]
  \centering
  \includegraphics[width=.45\textwidth]{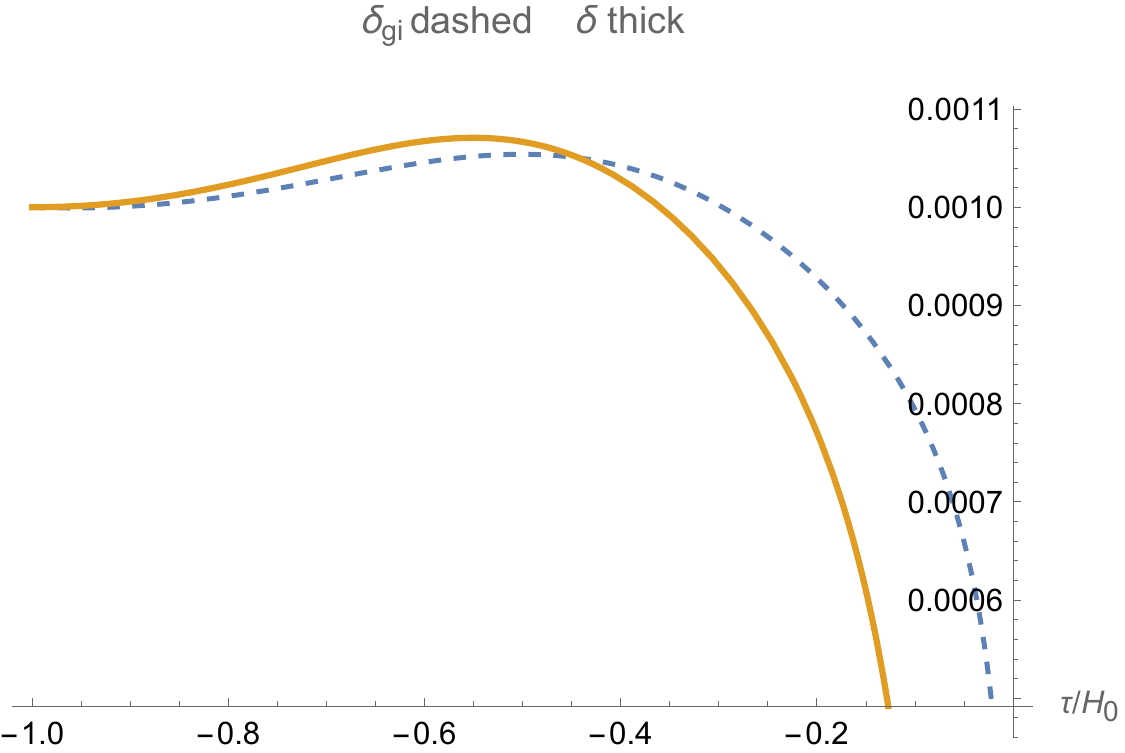}
    \includegraphics[width=.45\textwidth]{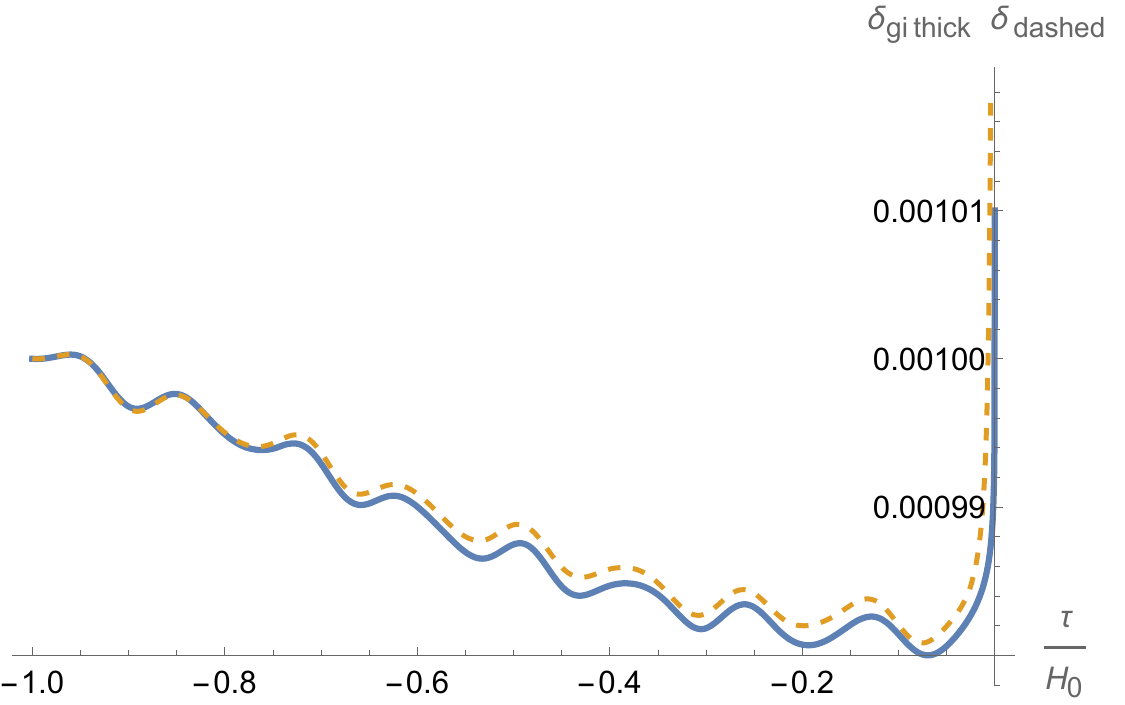}
    \caption{Matter overdensity numerical solutions. On the left hand
side the case  $k=10 \, H_0$ with $\delta_{gi}$ represented in a blue dashed line. On the right hand side the case
$k=10^2 \, H_0$, with $\delta_{gi}$ now in a solid blue line}.
\label{fig-delta1}
\end{figure}
On superhorizon scales, not only $\delta_m$ can be
hardly observed but also suffers from gauge ambiguities; as discussed
in appendix~\ref{app-gaugeinv} a much better quantity in this
respect is
\be
\delta_{gi}= \delta_m -3 \, {\cal H} \, v_m \; .
\ee
As expected, the form
of $\delta_{gi}$ and $\delta_m$ are very similar for subhorizon modes,
while are substantially different for superhorizon modes.
The numerical results are   shown in
figure \ref{fig-delta} and figure \ref{fig-delta1}.

In the LCDM model, during the cosmological constant domination,
regardless of the scale, both $\delta_m$ and $\delta_{gi}$ have no
growing mode and the same is true for $\Phi$ and $\Psi$; thus the advent of a late time dS phase marks the end
of structure formation. Things are different when the Universe is
dominated by the DCC
(\ref{ssol}): at the background level we still have a cosmological
constant except that  non-trivial perturbation exists. At
superhorizon scales, the power-law growth of the DCC scalar perturbations
$\pi_l$ and $\pi_0$ leads to the growth of the non relativistic
matter scalar velocity $v_m$ and matter contrast as shown in
figure \ref{fig-delta}. As depicted in figure \ref{fig-delta1}, the gauge
invariant matter contrast stays around 1 for large scales and
eventually  diverges as $1/a^{\frac{1}{2} \left(\sqrt{9-8
      c_2}+1\right)}$ near $\tau =0$, where a coordinates singularity
is encountered. At small scales the behaviour of matter contrast is rather
benign, and though stays small, is non-constant. As soon as the oscillatory regime of the dark energy scalar
perturbations sets in, it gets imprinted in the non-relativistic matter
contrast via eq. (\ref{eqdelta}), giving a series of dark energy
induced matter acoustic oscillations. As expected, no significant
difference is found between $\delta_m$ and $\delta_{gi}$ at small
scales. Once again, near $\tau =0$, both  $\delta_m$ and $\delta_{ gi}$
behave as $\frac{H_0}{k} \, a$.

\section{Gravitational Waves}
\label{gravwav}
Consider now perturbations corresponding to gravitational waves,
namely
\be
ds^2 = a^2 \left( \eta_{\mu \nu} \, dx^\mu  dx^\nu + \chi_{ij}   \,  dx^i
  dx^j \right) \, , \qquad \chi_{ij} \delta^{ij} = \de_j
\chi_{ij}=0 \, .
\ee
From the quadratic expansion of
\be
S= \frac{1}{8 \, \pi \, G} \int d^4x \sqrt{g} \left[ \frac{1}{2} \,  R + U(b, y, \chi, \tau_Y,
\tau_Z) \right] 
\ee
one arrives at the equation of motion in Fourier space for the spin 2
perturbations
\be
\chi_{ij}'' +2 \, {\cal H} \, \chi_{ij}' + \left(k^2 +  a^2 \, M_2 \right)
  \chi_{ij}=0 \, ,
 \label{teneq}
 \ee
 which is rather standard except for the additional mass term
 $M_2$. In dS the solution is a combination of Bessel functions of
 the form
 \be
 \chi_{ij}(\tau) = a^{-3/2} \, \epsilon_{ij} \left[ \chi_0 \, J_{\nu_T}(- k \tau) +
   \chi_1 \, Y_{\nu_T}( -k \tau) \right] \, , \qquad \nu_T= \left( 9 -
 4 \, c_2\right)^{1/2} \, .
   \ee
 For small scales ($ k \tau >> 1$) we have an oscillatory behaviour with
 a decreasing amplitude of the order $a^{-1}$, the very same behaviour
 as in LCDM. For large scales
 \be
 \chi_{ij} \sim  \epsilon_{ij}  \, a^{-3/2+ \nu_T/2} \, ,
 \ee
 and being  $c_2 > 0$, the amplitude grows logarithmically; a similar result was found in
an  inflationary context~\cite{Celoria:2020diz,Celoria:2021cxq}. In
contrast, when just a cosmological constant is present: $M_2=c_2=0$
and  the amplitude is constant at large scales. Such
a feature is difficult to test given the wavelength of the wave.

\section{Vector Modes}
\label{sect-vector}
Perturbations in the vector sector can be studied starting from the following metric:
\be
 ds^{2}=a^{2}(\eta_{\mu \nu} dx^{\mu}dx^{\nu}+2 B^{T}_{i}dx^{0}dx^{i}+2E_{ij}dx^{i}dx^{j})
 \label{pertVectMetric}
\ee
with 
\be
E_{ij}=\partial_{(i} E^{T}_{j)}=\frac{1}{2}(\partial_{i}
E^{T}_{j}+\partial_{j} E^{T}_{i}) \, , \qquad\partial^{i}
B^{T}_{i}=0 \, , \qquad\qquad\partial^{i} E^{T}_{i}=0 \, ;
\ee
while for the scalar fields we have
\be
    \Phi^{0}=\phi(t) \, , \qquad\qquad\qquad \Phi^{i}=x^{i}+\pi_T^{i} \, ,
    \qquad \de_i \pi^{Ti} =0 .
    \label{DEvectPert}
\ee
As in the scalar case, there is a gauge freedom also in the vector
sector and in this case, instead of fixing a gauge, we shall use gauge
invariant perturbations. As discussed in appendix \ref{app-gaugeinv},
the quantity $\pi^i_{gi}=\pi^{T}{}^i- E^T{}^i$ is gauge invariant and
transforms as a vector under the diagonal combination of internal and
spatial rotations and represents the vector physical
degrees of freedom in DCC medium. The dynamics of 
$\pi^i_{gi}$ can be obtained by a suitable combination of the conservation of the
medium's EMT and the Einstein equations. Indeed, 
from the $0i$ Einstein equations  is possible to express
$B^{T\,i}$ in terms of $\pi^{T\,i}$ and $\pi^{T\,i}_{\text{gi}}{}'$;
as a result, the relevant part of the medium's EMT turns into a
dynamical equation for $\pi^i_{gi}$ that, in the case $w=-1$
and by using the  definitions (\ref{cpar}), reads 
\be
\pi^i_{gi}{}'' +\frac{4
   \mathcal{H} \left(a^2 c_1 H_0^2+k^2\right)}{2 a^2 c_1
   H_0^2+k^2}\pi^i_{gi}{}' +\frac{c_2 \left(2 a^2 c_1
     H_0^2+k^2\right)}{c_1}\pi^i_{gi} =0 \, ,
\label{VectPropDE}
 \ee
where the scale factor $a$ is given by (\ref{ads}). The general expression valid for
any equation of state can be found in appendix \ref{app-gen}. For an ideal adiabatic
fluid  the classical
result of vorticity conservation (see for instance~\cite{Landau})
makes the dynamics of vector modes 
not very interesting.  Such a result stems from the large internal symmetries of
 the Lagrangian of an ideal adiabatic  fluid of the
 form $U(b,y)$~\cite{Celoria:2017bbh} is invariant under the following
 field dependent
 shift of $\Phi^0$ 
 \be
 \Phi^0 \to \Phi^0+f(\Phi^a) 
 \label{fluidsym}
 \ee
 and the volume-preserving diffeomorphisms (\ref{svolprd}).
 Vorticity conservation follows from the Noether theorem (for a
 recent discussion see~\cite{Ballesteros:2012kv}). As it is clear from (\ref{masses}), at the perturbative
 level,  the symmetries (\ref{fluidsym}) and (\ref{svolprd}) lead to $M_1=M_2=0$ or
 equivalently $c_1=c_2=0$. In our case the
 presence of the operator $\chi$ and $\tau_Y$ and $\tau_Z$ breaks
 (\ref{fluidsym}) and  vector modes do  propagate. 
% %
The dynamical equation for vector modes (\ref{VectPropDE}) can be
easily solved for superhorizon scales, namely when $k \tau \ll 1$,  and
it gives
\be
\pi _{\text{gi}}^i=q_1^i (- k \tau)^{(3- \sqrt{9 -8 c_2})/2}+ q_2^i (- k
\tau)^{(3+ \sqrt{9 -8 c_2})/2} \, ,
\ee
where $q_{1,2}^i$ are arbitrary transverse vectors in Fourier space
satisfying $k^i q_{1,2}^i=0$. In the opposite limit (small scales), $ k
\tau \gg 1$, one has
\be
\begin{split}
&\pi _{\text{gi}}^i= \sin \left(\frac{\sqrt{c_2}
   k \tau }{\sqrt{c_1}}\right) \left[ \frac{\sqrt{\frac{2}{\pi }}
   \sqrt[4]{c_1} \, \tilde q_2^i \left(c_2 k^2 \tau ^2-3 \, 
   c_1\right)}{c_2^{5/4}}+3 \sqrt{\frac{2}{\pi }}
\left(\frac{c_1}{c_2}\right){}^{3/4} k \, \tau  \, \tilde q_1^i\right] \\
&+\cos \left(\frac{\sqrt{c_2}
   k \, \tau }{\sqrt{c_1}}\right) \left[\frac{\sqrt{\frac{2}{\pi }}
   \sqrt[4]{c_1} \, \tilde q_1^i \left(c_2 \, k^2 \, \tau ^2-3
\,    c_1\right)}{c_2^{5/4}}-3 \, \sqrt{\frac{2}{\pi }}
   \left(\frac{c_1}{c_2}\right){}^{3/4} k\,  \tau \,  \tilde q_2^i \right]\, ,
\end{split}
 \ee
where again $\tilde q_{1,2}^i$ are arbitrary transverse vectors in Fourier
space. In both regimes there is no  growing mode.

\section{Conclusions}
\label{concl}
The $\Lambda$CDM model gives by far the most economical description of
dark energy in terms of a cosmological constant, no additional
degrees of freedom are needed. Our analysis shows that introducing
dynamics in the dark energy sector while keeping the equation of state
$w=-1$ is difficult. For instance, considering a
  k-essence theory based on a single scalar field with a perfect fluid
  EMT,  inevitably leads to a pathological dynamics for 
 perturbations when $w=-1$. One has  to move away from
 the description in terms of single scalar field minimally coupled
 with gravity. We have shown that it is possible to device a classical field theory that mimics 
  the very same equation of state of a cosmological constant but with
  a non-trivial dynamics by using  four  scalar fields minimally coupled with gravity, which gives an effective description of
  the most general non-dissipative medium: a combination of a solid
  and a superfluid. The EMT deviates from the one of a perfect fluid,
  in particular the existence of a non-vanishing anisotropic stress is
  a crucial requirement to avoid instabilities. 
We have also shown that the equation of state  $w=-1$ cannot be achieved  with a
positive definite Hamiltonian. Though the ``normal'' diagonal 
  scalar modes show a perfectly healthy oscillatory behaviour with
subluminal speeds of sound, the total Hamiltonian $H$ can be written
as the {\it difference} $H=H_{\omega_1} -H_{\omega_2}$    of two
harmonic oscillators, modulo a non-trivial canonical
transformation. The fear is
that adding a small interaction  when the energy of a system is  unbounded from bellow, 
 will  lead to instabilities. 
It turns out that, when the above dynamical cosmological constant model is minimally
coupled with gravity, no fast instability is
 found. At large scale, there is a mild power-like growth of
scalar perturbation.  The presence of small perturbation of the
non-relativistic 
matter contrast $\delta_m$ during dynamical cosmological constant
domination differs from 
the case of a cosmological constant  dominated Universe. While in the
LCDM model $\delta_m$ is constant during $\Lambda$ domination, in our case
at small scales $\delta_m$ shows small oscillations induced by the
fluctuations of the dynamical cosmological constant. Thus, phenomenologically
  the scalar sector is very  interesting. Also in the gravitational
  waves and vector sectors no instability is found. In particular,  at subhorizon
  scales the propagation of gravitational waves is the same as in LCDM
  and the speed of propagation is not altered; differences are found
  at superhorizon scales where the amplitude grows logarithmically
  instead of being constant.  Our analysis is based only on linear
  perturbation theory, it would be interesting to study the
  coupling with gravity  in the full non-linear regime and when
  self-interactions among the medium modes are considered. The presence of
  propagating vector modes and the difference between the two Bardeen
  scalar potentials induced by the  anisotropic stress in DCC EMT can influence the polarisation and the
  propagation of the CMB photons. We leave those
  matters for a future investigation. 
 
\appendix
  \section{k-Essence}
  \label{app-kess}
  In the present appendix we will study the dynamics of cosmological
  perturbation of a Universe dominated by k-essence scalar field
  $\Phi$ described by the Lagrangian $K(\Phi,X)$ where
  \be
  X = -\ha g^{\mu \nu} \de_\mu \Phi \de_\nu \Phi \, .
  \ee
Consider linear perturbations around a spatially FRW Universe; the perturbed scalar field and metric in the
$\zeta$ gauge read~\footnote{Notice that in the present appendix we
  do not use conformal time.}
\be
\begin{split}
&ds^2 = - \left(1+2 \,  \delta N \right) dt^2 +2 \, dt dx^i \, \de_i \psi + a^2 \,
e^{2 \, \zeta} \, \delta_{ij} dx^i dx^j \, ;\\
& \Phi \equiv \bar \Phi = \phi(t) \, .
\end{split} 
\ee
In the $\zeta$ gauge, part of the gauge  freedom is used to gauge away the
scalar field perturbation.
Expanding the total action, gravity plus scalar field, at the linear order in the
fields' perturbations gives the background equations of motion
\bea
&& 3 \, H^2 + 8 \pi G \left(\bar K + 2 \dot{\phi} \, \bar F_X \right)
=0 \, , \\
&& 3 \, H^2 + 2 \, \dot{H} + 8 \pi G \, \bar F =0 \, ,
\ena
where $H= \dot{a}/a$. In the case $w=-1$, namely de Sitter spacetime,
we have $H \equiv H_0$ is constant  and we get
\be
\bar K \equiv K(\phi, \bar X) = \text{constant} \, , \qquad \dot{\phi}
  \, \bar K_X =0 \, .
  \ee
In the quadric action it turns out that both $\delta N$ and $\psi$
have algebraic equations of motion and thus the only perturbation with
non-trivial dynamics is $\zeta$ and we get, up to total derivatives
terms, the following quadratic Lagrangian  in Fourier space
\be
L_\zeta^{(2)}= A_\zeta \, \dot{\zeta}^2 + B_\zeta \, \zeta^2  \, ,
\ee
with both $A_\zeta$ and $B_\zeta$ such that
\be
A_\zeta{}_{| \dot{\phi} =0}=0 \, , \qquad B_\zeta{}_{| \dot{\phi} =0}=0
\, .
\label{noshift}
\ee
If no shift symmetry is present, $\bar K$ can be constant only if
$\phi$ is constant and then $\dot{\phi}=0$; as a direct consequence of
(\ref{noshift}), $L_\zeta^{(2)}=0$ and the dynamics of the $\zeta$ is
pathological. If a shift symmetry is indeed present, then $H = H_0$ can
be obtained with $\dot{\phi}=$ constant and $\bar K_X=0$ with
\be
A_\zeta= \frac{2 \, a^3 \, K_{\text{XX}} \, \dot{\phi}^4}{H_0^2}
\ee
that is non-vanishing and can be easily made positive providing a
healthy kinetic term for $\zeta$. Unfortunately $B_\zeta$ is also
zero when  $\dot{\phi}$ is constant and $K$ does not depend on $\Phi$
and then the speed of sound of $\zeta$ is zero.

\section{Parameters}
  \label{lan}
 The parameters $\{M_A, \; A=0,1,2,4\}$ entering in (\ref{lang}) have
 the following expression in terms of the derivatives of $U$ in a
 spatially flat FRW background 
 \be
 \begin{split}
& M_0=\frac{\phi'{}^2 \left(U_{\chi \chi }+2 \, 
     U_{y \chi }+U_{yy}\right)}{2 \, a^2 } \, , \qquad
 M_1=-\frac{U_{\chi } \, \phi'}{a} \, , \qquad M_2 = -\frac{4
   \left(U_{\tau Y}+U_{\tau Z}\right)}{9 }, \\
&M_3= \frac{27 \, a^{-6} \, U_{bb}-8 \left(U_{\tau Y}+U_{\tau
      Z}\right)}{54 } \, , \qquad M_4 =\frac{\phi' \left \{-\left[a^3 \left(U_{\chi
        }+U_y\right)\right]+U_{b\chi}+U_{by}\right \}}{2\,  a^4 }  \, .
\end{split}
\label{masses}
 \ee
 In the case of Minkowski background one simply sets the scalar factor
 $a$ to 1 and $\phi'=1$.

 \section{Perturbed EMT}
 \label{app-EMT}
 The perturbation of the dark energy EMT at the linear order reads
\bea
&&
M^{-2} \, T_{00}^{1(DE)} = \nb\\
&&2  \left(a^2 M_4+\mathcal{H}^2-\mathcal{H}'\right) \Delta \pi _l+ \Psi 
   \left(6 \mathcal{H}^2-2 a^2 M_0\right)+6  \, \Phi \left(4 a^2
   M_4+\mathcal{H}^2-\mathcal{H}'\right)+\frac{2 a^2 M_0 \,
   \pi_0'}{\bar \phi '}  \\
 && M^{-2} \, T_{0i}^{1(DE)} = \de_i \pi _l' \left(a^2 M_1+2
   \mathcal{H}^2-2 \mathcal{H}'\right)-\frac{ a^2
   M_1}{\phi '}  \de_i \pi _0 \\
 &&  M^{-2} \, T_{ij}^{1(DE)} = \delta_{ij} \left[2 a^2 \left( M_3 \,
     \Delta \pi _l-\frac{M_4 }{\phi '} \pi _0' \right)+2 a^2 M_4
   \Psi+2 \Phi \left(a^2 \left(M_2-3
       M_3\right)+\mathcal{H}^2+2 \mathcal{H}'\right) \right]  \nb \\
 &&+ 2 \, M_2 \, a^2 \, \de_i \de_j \pi_l \, .
\ena
The above expression holds for any equation of state; when
$w=-1$ and the background is a portion of dS space, $T_{\mu
  \nu}^{1(DE)} $ reduces to
\bea
&& M^{-2} \, T_{00}^{1(DE)}{}_{w=-1} = 2 \, a^2 \, M_0 \, \Delta \pi _l +6 a^2 M_0 \,
\Phi+\Psi \left(6 \mathcal{H}^2-2 a^2 M_0\right)+\frac{2 M_0 \pi _0'}{a^2} \, ; \\
&& M^{-2} \, T_{0i}^{1(DE)}{}_{w=-1}  = M_1 \left(a^2  \de_i \pi _l' - a^{-2}
  \, \de_i \pi _0 \right) \, ;\\
&&  M^{-2} \, T_{ij}^{1(DE)}{}_{w=-1} =\delta_{ij} \left[ 2 \, \Phi  \left(3 \mathcal{H}^2-2 a^2 M_0\right)-\frac{2 M_0 \pi
   _0'}{a^2}-2\left(\frac{1}{3} a^2 k^2 \left(3 M_0+M_2\right)-2 a^4 M_0
   M_2\right)  \Delta \pi _l \right] \nb \\
&&+ 2 \, M_2 \, a^2 \, \de_i \de_j \pi_l \, 
\ena
and cannot be described as a perturbed perfect fluid. Notice that the
EMT of the dark energy is linear in the perturbations as a consequence
of the spontaneous symmetry breaking of translations.

\section{Gauge Invariant Perturbations}
\label{app-gaugeinv}
Let us consider the metric
\be
ds^2= a^2 \left[ -(1+ 2 \, A ) dt^2+ 2 \, B_i \, dx^i d t + \left(
    \delta_{ij} 2 \, E_{ij}  \right) dx^i dx^j \right] \equiv \left(
  a^2 \, \eta_{\mu \nu} + h_{\mu \nu} \right) dx^\mu dx^\nu \, ;
\ee
with
\be
\begin{split}
&B_i =\de_i B + B_i^T \, \qquad E_{ij} = C \delta_{ij} + \left(\de_i
  \de_j - \frac{1}{3} \nabla
\right) E +\ha \left( \de_i E_j^T+ \de_j E_i^T \right) +\chi_{ij} \, ; \\
& \de_i B_i^T= \de_i E_i^T =0 \, , \qquad \de_j \chi_{ij}= \delta_{ij}
\chi_{ij} =0 \,,  \qquad \nabla = \delta_{ij} \de_i \de_j \, .
\end{split}
\ee
The linear perturbations can be decomposed into: 4 scalars ($A$, $B$,
$C$, $E$), 2 transverse vector ($B_i^T$, $E_i^T$) and a transverse
and traceless tensor ($\chi_{ij}$) according to 3D rotational group,
for a total of $4+4+2=10$ components. Under an infinitesimal
coordinate transformation: $x^\mu \to x^\prime{}^\mu=x^\mu -\xi^\mu$, the
metric transforms at the linear order as 
\be
\delta g_{\mu \nu}(x) = g_{\mu \nu}'(x)- g_{\mu \nu}(x) \equiv a^2 \,
\delta h_{\mu \nu}(x) = \xi^\alpha \de_\alpha \left(a^2 \, \eta_{\mu
    \nu} \right) + a^2 \left( \eta_{\mu \alpha} \, \de_\nu \xi^\alpha +\eta_{\nu \alpha} \, \de_\mu \xi^\alpha 
  \right)  \, .
\ee
%%%%
Decomposing $\xi^i$ as $\xi^i = \de_i \xi + \xi^T{}^i$, with $\de_i
\xi^T{}^i=0$, one gets the following transformation properties for the
perturbations
\bea
&& \delta C = {\cal H} \, \xi^0 + \frac{1}{3} \nabla \xi \, , \qquad \delta A
= {\cal H} \, \xi^0 + \de_t \xi^0 \, , \qquad \delta B= \de_t \xi - \xi^0 \,
, \qquad \delta E = \xi \\
&& \delta B^T{}^i = \de_t \xi^T{}^i \, , \qquad  \delta E^T{}^i =
\xi^T{}^i \, .\\
&& \delta \chi_{ij} =0 \, .
\ea
Because of the redundancy induced by $\xi^0$ and $\xi^T{}^i$, only two
scalars and a single transverse vector are physical. The identification of
such physical perturbation can be made  by identifying special ``gauge invariant'' combination of the
perturbations that remain unchanged under an infinitesimal coordinate
transformation. A possible choice is given by the Bardeen potentials~\cite{Bardeen:1980kt}
\be
\Phi=-C+ {\cal H}(\de_t E-B)+\frac{1}{3} \nabla E \, , \qquad \Psi=
A+\de_t \left(B-\de_t B \right)+ {\cal H} \left(B-\de_t E \right) \, ;
\ee
with $\delta \Phi=\delta \Psi=0$. In the Newtonian gauge used in
(\ref{gpertnewt}) $\xi^0$ and $\xi$ are chosen in  such a way that
$E_{\text{Newt}}=B_{\text{Newt}}=0$, then
\be
\Psi=\Psi_{\text{Newt}}=A \, , \qquad \Phi=\Phi_{\text{Newt}}=-C \, .
\ee
For a  scalar field $f= \bar f + f_1 + \cdots$ with a non-vanishing
background value, its linear perturbation transforms as
\be
\delta f_1 = \xi^\mu \de_\mu \bar f \, .
\label{scaltransf}
\ee
The three scalar fields $\{\Phi^1, \, \Phi^2 , \, \Phi^2  \}$
transform as a vector under ``internal'' rotations and  once they get the
background value  (\ref{backval}) they also transform as a vector
under spatial rotation, namely  
\be
\begin{split}
& SO(3)_I: \quad \Phi^a \to \Phi^\prime{}^a= {\cal R}^a_b \, \Phi^b \,
, \qquad x^a \to x^a \, ;\\
& SO(3)_S: \quad \Phi^a\to \Phi^a \, ,
, \qquad x^a \to x^\prime{}^a= {\cal R}^a_b \, x^b \, , \qquad
a,b=1,2,3 \, .
\end{split}
\ee
Thus the original symmetry of the action $SO(3)_I \times SO(3)_S$  is broken down a
diagonal $SO(3)$ that leaves invariant the background configuration
$x^a$. We can decompose the perturbations $\pi^a$ in a vector and a scalar
part according to
\be
\pi^i = \pi^T{}^i + \de_i \pi_l \, , \qquad \de_i \pi^T{}^i=0 \, .
\ee
From (\ref{scaltransf}) we deduce that
\be
\begin{split}
&\delta \pi_l = \xi \, , \qquad \delta \pi^T{}^i =\xi^i{}^T \, , \qquad
i=1,2,3 \, . \\
& \delta \pi_0 =\xi^0 \, \phi' \, .
\end{split}
\ee
As a result one can construct the gauge invariant generalisation of
$\pi_l$ and $\pi_0$, namely
\be
\pi_{0\text{gi}} =\pi_0 - \frac{1}{\phi'} \, (\de_t E-B) \, , \qquad
\pi_{l\text{gi}} =\pi_l -E \, .
\ee
Thus, in the Newtonian gauge,  $\pi_l$ and $\pi_0$ and their gauge
invariant generalisation coincide and do not suffer from
gauge ambiguities. 
For structure formation an important quantity is the
non-relativistic matter overdensity $\delta_m={\bar \rho}^{-1} \delta
\rho_m$ which is not gauge invariant. In general, the density $\rho_w$
of a perfect fluid with equation of state $w$, being a scalar, is
such that the density perturbation $\rho_w^{(1)}$ transforms as
\be
\delta \rho_w^{(1)} =\bar \rho_w \, \xi^0 \, .
\label{trrho}
\ee
Considering the corresponding  4-velocity $u^\mu$ of the fluid one has
\be
\begin{split}
&u^\mu = \bar u^\mu + u^{(1)}{}^\mu  \, , \qquad \bar u^\mu =
\delta^\mu_0 \, a^{-1} \, , \\
&u^{(1)}{}^\mu = -\delta^\mu_0 \, A
+ \delta^\mu_i \left({v_w}^T{}^i_ + \de_i v_w \right) \, , \qquad \de_i
{v_w}^T{}^i =0 \, ;
\end{split}
\ee
with
\be
\delta v_m = -\de_t \xi \, , \qquad \delta {v_m}^T{}^i= - \xi^T{}^i
\, .
\label{trv}
\ee
Exploiting (\ref{trrho}) and (\ref{trv}) one can easily construct the
following gauge invariant generalisation of the  density perturbation 
\be
\delta_{w\text{gi}}= \frac{\rho_w^{(1)}}{\bar \rho_w} -3 (1+w) {\cal H}
\left(B+v_w \right)  \, .
\label{deltagi}
\ee
For vector modes, one can form out of the metric perturbations a single
independent gauge invariant combination
\be
E_{Tgi}^i=  E^{Ti}- \de_t B^{Ti} \, .
\ee
For the vector part of $\pi^a$ one can easily
define the following gauge invariant vector perturbation
\be
\pi^i_{gi} = \pi^{T}{}^i- E^T{}^i \, .
\label{pivecgi}
\ee

\section{Perturbations: generic equation of state}
\label{app-gen}

In the main text we have studied the special case $w=-1$ which leads
at the background level to de Sitter spacetime during dark energy
domination. In the appendix we report the general expressions valid for
any equation of state.

For a generic equation of state the relations
(\ref{Mrel}) do not hold anymore and it is convenient to define
\be
c_b^2= - \frac{M_4}{M_0} \, ;
\ee
when $w=-1$,  $c_b^2=-1$ and then from (\ref{phieq}) one gets by
integration $\phi^\prime=a^4$; moreover
\be
w = -\frac{2 \mathcal{H}'}{3 \mathcal{H}^2}-\frac{1}{3} \, ,
\ee
and in general for $w \neq -1$, $w$ can be time-dependent.

Let us consider scalar modes first. From the spatial part of the
perturbed Einstein equations one expresses $\Psi$ and $\Phi$ in terms of
the $\pi_0$ and $\pi_l$; eq. (\ref{anieq}) holds for any equation of
state while from the 00 component of perturbed Einstein equations one
finds the generalisation of (\ref{Phieq}), namely
\be
\begin{split}
&\Phi=\frac{\pi _l \left[k^2 \left(\mathcal{H}^2-\mathcal{H}'\right)-2 a^4 M_0
   M_2\right]}{Q}-\frac{3 \mathcal{H} \pi _l' \left(a^2 M_1+2
   \mathcal{H}^2-2 \mathcal{H}'\right)}{2 Q}+\frac{3 \pi _0 a^2 M_1
 \mathcal{H}}{2 Q \phi '}-\frac{a^2 M_0 \pi _0'}{Q \phi '} \, ,\\
& Q=k^2-\left[a^2 M_0 \left(3 c_b^2+1\right)\right]+3 \mathcal{H}^2-3
   \mathcal{H}' \, ;
\end{split}
\ee
The dynamical equations for $\pi_l$ and $\pi_0$ that
follow from the EMT conservation are more involved
\be
\begin{split}
&\pi_l''+\pi _l' \left \{\frac{a^2 \left[\mathcal{H} \left(3 M_1-2
   M_2+6 M_3 -6 c_b^4 M_0\right)+M_1'\right]}{a^2 M_1+2 \mathcal{H}^2-2
   \mathcal{H}'}+\frac{\mathcal{H} \left[a^2 \left(M_0+3 M_2-9
       M_3\right)-k^2\right]}{Q}+2 \mathcal{H}\right \} + \\
&\frac{a^2 \left \{M_0 \left[3 a^2 \, c_b^4 M_0+a^2 \left(M_2-3
   M_3\right)-c_b^2 \left(k^2+3 \mathcal{H}^2-3
   \mathcal{H}'\right)+\mathcal{H}^2-\mathcal{H}'\right]-M_1 \, Q\right \}}{Q
   \phi ' \left(a^2 M_1+2 \mathcal{H}^2-2 \mathcal{H}'\right)}\pi_0'+\\
 &\frac{{\cal F}_4 \, k^4 +{\cal F}_2 \, k^2 +{\cal F}_0} {Q \left(a^2 M_1+2 \mathcal{H}^2-2 \mathcal{H}'\right)}\pi_l 
 + \frac{a^2 \left( {\cal G}_2 \, k^2 + {\cal G}_0 \right)}{Q \phi ' \left(a^2 M_1+2 \mathcal{H}^2-2
     \mathcal{H}'\right)} \pi _0  =0  \, ; \\
 & \pi _0''+\frac{3 a^2 M_0^2 \left(3 \mathcal{H} c_b^4+\mathcal{H}
   c_b^2-c_b'{}^2\right)-2 M_0 \mathcal{H} \left(3
   c_b^2+1\right) \left(k^2+3 \mathcal{H}^2-3 \mathcal{H}'\right)-M_0'
   \left(k^2+3 \mathcal{H}^2-3 \mathcal{H}'\right)}{M_0 Q} \pi _0' + \\ 
 &\frac{ \phi ' \left({\cal A}_4 \, k^4 +{\cal A}_2\,  k^2+ {\cal A}_0
   \right)}{2 M_0 Q} \pi _l' -\frac{M_1\left(  k^4 +  {\cal B}_2 \,
     k^2+ {\cal B}_0 \right) }{2 M_0 Q}\pi _0+ \frac{\phi '\left({\cal
       C}_4 \, k^4 +{\cal C}_2 \, k^2 +{\cal C}_0
 \right) }{M_0 Q} \pi _l  =0 \, ,
\end{split}
\ee
where
\be
\begin{split}
&  {\cal F}_4=2 a^2 \left(M_2-M_3\right) \, ,\\
  & {\cal F}_2=8 a^2 M_2 \left(\mathcal{H}^2-\mathcal{H}'\right)-2
   \left(\mathcal{H}^2-\mathcal{H}'\right)^2  +2 a^2 M_0 \left \{ 2 \,
     c_b^2 (\mathcal{H}^2-
   \mathcal{H}')-\left[a^2 \left(c_b^4 M_0+\left(4 c_b^2+1\right)
   M_2-M_3\right)\right] \right \} , \\
   &{\cal F}_0=12 a^6 M_0^2 M_2 c_b^4+12 a^2 M_2
   \left(\mathcal{H}^2-\mathcal{H}'\right)^2+4 a^4 M_0 M_2 \left[a^2
   \left(M_2-3 M_3\right)+6 c_b^2
   \left(\mathcal{H}'-\mathcal{H}^2\right)\right] 
\end{split}
\ee
and
\be
\begin{split}
  &  {\cal G}_2=-3 \left(1+ c_b^2\right) M_1 \mathcal{H}\\
   &  {\cal G}_0=3 M_1 \mathcal{H} \left[a^2 \left(3 \, c_b^4+5 \, c_b^2+1\right)
   M_0-a^2 \left(M_2-3 M_3\right)-\left(3 \, c_b^2+4\right)
   \left(\mathcal{H}^2-\mathcal{H}'\right)\right]-Q M_1' \, ,\\
 &{\cal A}_4 =2 M_0 c_b^2+M_1 , \\
 &{\cal A}_2 = -2 M_0 \left(a^2 \left(M_1 \left(3 c_b^2+1\right)-2
   M_2\right)+\mathcal{H}^2-\mathcal{H}'\right)-2 a^2 M_0^2 c_b^2 \left(3
   c_b^2+1\right)+3 M_1 \left(\mathcal{H}^2-\mathcal{H}'\right), \\
 &{\cal A}_0 =3 M_0 \Big[9 \mathcal{H}^2 c_b^4 \left(a^2 M_1+2 \mathcal{H}^2-2
   \mathcal{H}'\right)+3 c_b^2 \left(2 \mathcal{H}^2+\mathcal{H}'\right)
   \left(a^2 M_1+2 \mathcal{H}^2-2 \mathcal{H}'\right)+a^2 M_1 \left(3
   \mathcal{H} c_b'{}^2+\mathcal{H}^2+\mathcal{H}'\right)\\
 &+2
   \left(\mathcal{H}^2-\mathcal{H}'\right) \left(2 a^2 M_2+3 \mathcal{H}
     \left(c_b'\right){}^2+\mathcal{H}^2+\mathcal{H}'\right)\Big ]
 +3
   \mathcal{H} \left(3 c_b^2+1\right) M_0' \left(a^2 M_1+2 \mathcal{H}^2-2
     \mathcal{H}'\right) , \\
  & {\cal B}_2 = 3 \left(\mathcal{H}^2-\mathcal{H}'\right)-2 a^2 M_0
  \left(3 c_b^2+1\right) , \\
 & {\cal B}_0= a^2 \left(M_0^2 \left(3 a c_b^2+a\right){}^2+3 M_0 \left[\left(3
   c_b^2+1\right) \left(3 \mathcal{H}^2
   c_b^2+\mathcal{H}^2+\mathcal{H}'\right)+3 \mathcal{H}
   c_b'{}^2 \right]+3 \mathcal{H} \left(3 c_b^2+1\right)
 M_0'\right) , \\
&{\cal C}_4 = M_0 \left[3 \mathcal{H} \left(c_b^2+1\right)
  c_b^2+c_b'{}^2\right]+c_b^2 M_0' , \\
&{\cal C}_2 =-a^2 M_0^2 \left(3 \mathcal{H} c_b^4+\mathcal{H}
   c_b^2+c_b'{}^2\right)+2 M_0 \left(4 a^2 M_2 \mathcal{H}+a^2
   M_2'-\mathcal{H}^3+\mathcal{H} \mathcal{H}'\right)+M_0' \left(2 a^2
   M_2-\mathcal{H}^2+\mathcal{H}'\right), \\
 & {\cal C}_0 =2 a^2 \Big \{ a^2 M_0^2 \left[M_2 \left(\mathcal{H} \left(9 c_b^4-3
   c_b^2-2\right)+3 c_b'{}^2\right)-\left(3 c_b^2+1\right)
M_2'\right]\\
&+3 M_0 \left(\mathcal{H}^2-\mathcal{H}'\right) \left(4 M_2
   \mathcal{H}+M_2'\right)+3 M_2 M_0'
   \left(\mathcal{H}^2-\mathcal{H}'\right)\Big \} \, .
 \end{split}
\ee
For vector modes, the dynamical equations  can be obtained from the
$0i$ component
of Einstein equations and from the conservation of the EMT; the result is
\be
\begin{split}
&\pi_{gi}^i{}''+ 22 a^2 M_2 \left(\frac{k^2}{2 a^2 M_1+4 \mathcal{H}^2-4
    \mathcal{H}'}+1\right) \pi_{gi}^i \\
&+\Big \{4 a^4 M_1 \mathcal{H} \left(-3 M_0 c_b^2+M_1-2 M_2+3 M_3\right)+a^2
   \big [-6 M_0 \mathcal{H} c_b^2 \left(k^2+4 \mathcal{H}^2-4
       \mathcal{H}'\right) \\
     & +4 M_1 \left[\mathcal{H} \left(k^2+3 \mathcal{H}^2-5
     \mathcal{H}'\right)+\mathcal{H}''\right] \\
 & -2 k^2 M_2 \mathcal{H}+6 k^2 M_3
   \mathcal{H}+k^2 M_1'-8 M_2 \mathcal{H}^3+24 M_3 \mathcal{H}^3+8
   \left(M_2-3 M_3\right) \mathcal{H} \mathcal{H}' \big ]\\
 &+2
   \left(\mathcal{H}^2-\mathcal{H}'\right) \left(\mathcal{H} \left(k^2-12
   \mathcal{H}'\right)+4 \mathcal{H}^3+4 \mathcal{H}''\right)  \Big \}\, \pi_{gi}^i{}' =0 \, .
\end{split}
\ee
For what concerns tensor modes, (\ref{teneq}) holds for any equation of state.

\bibliography{biblio-DE}

\end{document}